\documentclass[prd,%
reprint, %% this has to be changed to reprint to match more closely the paper's appearance
superscriptaddress,
showpacs,
nofootinbib,
 amsmath,amssymb,
 aps,
floats,
floatfix,
twocolumn]{revtex4-1}

\usepackage{graphicx}% Include figure files
\usepackage{dcolumn}% Align table columns on decimal point
\usepackage{bm}% bold math
\usepackage{color}
\usepackage{url}
\usepackage{soul}
\newcommand{\PR}[1]{\ensuremath{\left[#1\right]}} % parenteses rectos do tamanho adequado
\newcommand{\PC}[1]{\ensuremath{\left(#1\right)}} % parenteses curvos do tamanho adequado
\usepackage{hyperref}% add hypertext capabilities
\begin{document}

%\preprint{APS/123-QED}

\title{Constraints on decaying early modified gravity from cosmological observations}

\author{Nelson A. Lima}
 \email{ndal@roe.ac.uk}
 \author{Vanessa Smer-Barreto}
 \email{vsm@roe.ac.uk}
 \author{Lucas Lombriser}
 \email{llo@roe.ac.uk}

%\thanks{}
\affiliation{Institute for Astronomy, University of Edinburgh, Royal Observatory, Blackford Hill, Edinburgh, EH9 3HJ, UK}

\renewcommand{\abstractname}{Abstract}
\begin{abstract} Most of the information on our cosmos stems from either late-time observations or the imprint of early-time inhomogeneities on the cosmic microwave background. We explore to what extent early modifications of gravity, which become significant after recombination but then decay towards the present, can be constrained by current cosmological observations. For the evolution of the gravitational modification, we adopt the decaying mode of a hybrid-metric Palatini $f(\mathcal{R})$ gravity model which is designed to reproduce the standard cosmological background expansion history and due to the decay of the modification is naturally compatible with Solar-System tests. We embed the model in the effective field theory description of Horndeski scalar-tensor gravity with an early-time decoupling of the gravitational modification. Since the quasistatic approximation for the perturbations in the model breaks down at high redshifts, where modifications remain relevant, we introduce a computationally efficient correction to describe the evolution of the scalar field fluctuation in this regime. We compare the decaying early-time modification against geometric probes and recent Planck measurements and find no evidence for such effects in the observations. Current data constrains the scalar field value at $|f_{\mathcal{R}}(z=z_{\rm on})| \lesssim 10^{-2}$ for modifications introduced at redshifts $z_{\rm on}\sim(500-1000)$ with present-day value $|f_{\mathcal{R}0}|\lesssim10^{-8}$. Finally, we comment on constraints that will be achievable with future 21~cm surveys and gravitational wave experiments.
\end{abstract}

\pacs{98.80.-k, 95.36.+x, 04.50.Kd \hfill \today}% PACS, the Physics and Astronomy
\maketitle

%\tableofcontents

\section{\label{Int}Introduction}

%\lucas{Can you update the references. Some have been published in the meantime.}
%\vanessa{Lovelock's theorem is proof [\nelson{shows/proves, perhaps?}] that Einstein's equations are the only second-order local equations of motion for a metric derivable from the gravitational action in four dimension.
%Therefore, any modifications of general relativity (GR) should add one or more of the following elements: extra degrees of freedom, higher derivatives, extra dimensions or imply non-locality~\cite{koyama:15}.}
%\lucas{While I like the intro, these are not exactly the conditions that go into the Lovelock Thm. The 1969 paper is not so clear about this. Check out the 1970 and 1972 follow-up papers instead. How about \ldots}
%According to Lovelock's theorem, any gravitational action that differs from the one originally devised by Einstein will imply either new degrees of freedom, higher order equations of motion or extra dimensions.
%In the presence of extra gravitational degrees of freedom or extra dimensions evades Lovelock's theorem and modifies Einstein's gravitational field equations.

Lovelock's theorem on the uniqueness of Einstein's gravitational field equations does not apply in the presence of higher than second-order derivatives of the metric, additional gravitational degrees of freedom, extra dimensions, or other unconventional properties that a more fundamental theory of gravity may be endowed with.
Such a theory could give rise to new phenomenological aspects at different scales and epochs in time that may potentially be observable and can be tested using existent modified gravity theories.
%as toy models.

Einstein's Theory of General Relativity (GR) has been well tested in the Solar System, where, however, potential large-scale deviations may be suppressed due to screening effects~\cite{reviewall, koyama:15, joyce:16}.
%are tightly constrained on the smallest scales where we need to recover standard GR, with Solar system tests providing the tightest constraints we currently have on modified theories of gravity.
%
There is now a complementary effort in obtaining competitive constraints on larger scales, with a surge of surveys
%in the next decade
that will significantly improve our knowledge
%of the Universe on cosmological scales
of the cosmological regime, such as the Dark Energy Survey (DES) \cite{des}, the extended Baryon Oscillation Spectroscopic Survey (eBOSS) \cite{eboss} and the Euclid survey \cite{euclid} (for a review on cosmological tests of gravity see \cite{koyama:15}).

%Modified gravity theories (MGT) were introduced to explain our Universe's accelerated expansion~\cite{accel1,accel2,accel3,accel4} as an alternative to the standard model of cosmology $\Lambda$CDM, which suffers from a well known fine-tuning problem (for a review see \cite{lambdareview}).
Much of the interest in modified gravity theories has arisen in the search for alternative explanations for the observed late-time accelerated expansion of our Universe~\cite{accel1,accel2,accel3,accel4},
%that could
possibly avoiding the fine-tuning problem of the cosmological constant $\Lambda$ adopted in the standard model of cosmology $\Lambda$CDM (for reviews on modified gravity, the cosmological constant, and dark energy see \cite{lambdareview,reviewall,koyama:15,joyce:16}).
However, in Ref.~\cite{lombriser:16} it has recently been shown that scalar-tensor theories of gravity such as Brans-Dicke~\cite{bd}, Galileon~\cite{Gall}, and $f(R)$ gravity~\cite{frreview}, or any other models embedded in the Horndeski action~\cite{horndeski:74} cannot yield an observationally compatible self-acceleration effect due to modified gravity that its genuinely different from $\Lambda$ or dark energy, unless the cosmological speed of gravitational waves differs substantially from the speed of light.
%(for reviews on MGT and DE see~\cite{reviewall,joyce:16}).
%
While such a deviation is unlikely~\cite{lombriser:15b}, modified gravity theories are nevertheless relevant to test gravity and understand how it acts across all scales and epochs in cosmic time.

However, given the original interest in cosmic acceleration, the study of modified gravity has predominantly focused on late-time effects with a recovery of GR at high redshifts.
Hence, early-time modifications have, so far, evaded a thorough investigation and when they have been studied (e.g.~\cite{lombriser:11,brax:13}), their effects at early times have not been clearly separated from their late-time effects.
The missing analysis of early-time modifications and their impact on cosmological observables constitutes a gap in our current understanding of the gravitational processes at work and we lack a consistent quantitative analysis of the constraining power current (and future) cosmological surveys have over this regime of gravity.
Generally, the assumption of GR at early times without a test against alternatives is a strong extrapolation from its exclusive validity in the late-time time Solar-System region (or even from late-time cosmology).
This investigation is also important to quantify the improvement on our current understanding of the cosmological model that can be expected with future surveys such as $21$-cm intensity mapping (see, e.g.,~\cite{21cmdark} for expected dark energy constraints), the use of gravitational waves as standard sirens at high redshifts, or constraints from surveys such as the Square Kilometer Array (SKA) on the horizon~\cite{ska}.

In this paper, we explore to what extent modifications of gravity that may arise after recombination and decay towards the present can be constrained with current cosmological observations that stem either from their impact on the late-time large scale-structure or changes in the imprint of early-time inhomogeneities on the cosmic microwave background. For this purpose, we adopt the decaying mode of a hybrid metric-Palatini gravity model, which enables us to separate early- from late-time effects. We then compare the constraining power of future $21$-cm intensity mapping and standard sirens to these current constraints.

%We also explore beyond these two limiting regimes, and look at the constraining capabilities of observables based on $21$-cm intensity mapping and the use of gravitational waves as standard sirens. With surveys such as the Square Kilometer Array (SKA) on the horizon \cite{ska}, it is important to quantify the improvement one can have from these future surveys (also see \cite{21cmdark} for expected constraints on the dark energy equation of state from $21$-cm observations).
The outline of the paper is as follows. In Sec.~\ref{I}, we introduce and discuss the
%the model we will investigate in order to discuss the concept of
early-time decaying modified gravity model adopted for our analysis. In Sec.~\ref{perturbations}, we reproduce its linearly perturbed modified Einstein equations in the Newtonian gauge. We explicitly show how the breakdown of the quasistatic approximation for the evolution of the scalar field fluctuation occurs at high redshifts.
This failure motivates an analytic correction to the quasistatic approximation to accurately describe the evolution of the slip between the metric potentials in this high-curvature regime.
In Secs.~\ref{sec:decoupling} and \ref{coneft}, we describe an embedding of this gravitational modification in the effective field theory (EFT) of Horndeski scalar-tensor gravity (reviewed in Ref.~\cite{gleyzes:14}) with a post-recombination high-redshift decoupling of the modification to comply with stringent high-curvature constraints from the cosmic microwave background (CMB).
In Sec.~\ref{II}, we infer constraints on the early-time decaying modified gravity model using current cosmological observations. Lastly, in Sec.~\ref{III} we conclude with some final thoughts and remarks, also providing an outlook for future cosmological constraints on the model.
For completeness, in the appendices we provide details on our numerical computations and approximations adopted to describe oscillations in the scalar field fluctuations.

\section{A Decaying Early Modification of Gravity} \label{sec:model}

The main purpose of this work is to explore constraints on early modified gravity, with modifications from GR arising at high redshifts and being suppressed as we approach the present time.
We start by describing the general dynamics of the test model we will embed in Horndeski theory: the hybrid metric-Palatini $f(\mathcal{R})$ gravity, where the metric and the connection are considered as independent variables. 

Note that while metric $f(R)$ theory, where the connection is not independent, is much more frequently adopted as toy model to study modifications of gravity, and also possesses a decaying mode \cite{metricfrpert}, it naturally predicts a 4/3 enhancement of the effective gravitational coupling in unscreened observables at late times and small scales.
%do not provide a truly clean modification of gravity at the smallest scales. It is well known that $f(R)$ theories predict \nelson {an enhancement of the effective gravitational constant at linear order} , thus needing a screening mechanism to pass the stringent solar-system tests\cite{khoury:03,hu:07,brax:08,lombriser:14a}.
There always exists a small enough object in a late-time, low-density environment that is not screened and hence exhibits a modified gravity effect that could potentially be used to constrain the modification, for instance, a dwarf galaxy in a void \cite{dwarf_galaxies}.
Similarly, an upweighting of low-density regions in statistical observations of the large-scale structure can be used to effectively unscreen the modifications~\cite{lombriser:15}.
Hybrid metric-Palatini gravity evades these constraints as the unscreened effective gravitational coupling itself tends to the Newtonian value at late times (this argument will be explained in more detail in Sec.~\ref{subapp}). 

\subsection{\label{I}Hybrid Metric-Palatini Gravity}
The four-dimensional action describing the hybrid metric-Palatini gravity is given by~\cite{main1}
\begin{equation}{\label{action}}
 S = \frac{1}{2\kappa^2} \int d^4 x \sqrt{-g}\PR{R + f(\mathcal{R})} + S_{m} \,,
\end{equation}
\noindent where $\kappa^2 = 8 \pi G$ and we set $c=1$. $S_m$ is the standard matter action, $R$ is the metric
Ricci scalar and $\mathcal{R} = g^{\mu \nu} \mathcal{R}_{\mu \nu}$ is the Palatini curvature. The latter is defined in terms of the metric elements, $g^{\mu \nu}$, and a torsion-less independent connection, $\hat{\Gamma}$, through
\begin{equation}
\mathcal{R} \equiv g^{\mu \nu}\PC{\hat{\Gamma}^{\alpha}_{\mu \nu, \alpha} - \hat{\Gamma}^{\alpha}_{\mu \alpha,\nu} + \hat{\Gamma}^{\alpha}_{\alpha \lambda} \hat{\Gamma}^{\lambda}_{\mu \nu} - \hat{\Gamma}^{\alpha}_{\mu \lambda} \hat{\Gamma}^{\lambda}_{\alpha \nu}} \,.
\end{equation}
For a statistically spatially homogeneous and isotropic universe with Friedmann-Robertson-Walker (FRW) metric,
$ds^2 = - dt^2 + a^{2}(t)d\vec{x}^2$,
the modified Einstein equations and the dynamical hybrid-metric scalar field equation yield the modified Friedmann equations and background scalar field equation~\cite{main1,main2}:
\begin{eqnarray}
 3H^{2} & = & \frac{1}{1 + f_{\mathcal{R}}} \PR{\kappa^{2} \rho - 3 H \dot{f_{\mathcal{R}}} - \frac{3\dot{f_{\mathcal{R}}}^2}{4f_{\mathcal{R}}} + \frac{\mathcal{R}f_{\mathcal{R}}- f(\mathcal{R})}{2}} \,, \nonumber\\ {\label{hyb1}}\\
 2\dot{H} & = & \frac{1}{1 + f_{\mathcal{R}}} \PR{ -\kappa^{2}\PC{\rho + p} + H\dot{f_{\mathcal{R}}} - \ddot{f_{\mathcal{R}}} + \frac{3 \dot{f_{\mathcal{R}}}^{2}
 }{2f_{\mathcal{R}}}} \,, {\label{hyb2}} \\ 
\ddot{\mathcal{R}} & = & -\frac{1}{f_{\mathcal{RR}}} \PR{ \dot{{\mathcal{R}}}^2\left(f_{\mathcal{RRR}} - \frac{{f_{\mathcal{RR}}}^2}{2f_{\mathcal{R}}}\right) +3H\dot{\mathcal{R}}f_{\mathcal{RR}} \right.\nonumber \\
& & \left. + \frac{f_{\mathcal{R}}}{3}\PR{\mathcal{R}(f_{\mathcal{R}} -1) -2f(\mathcal{R})} -\kappa^2\frac{f_{\mathcal{R}}}{3}T} \,, {\label{hyb3}}
\end{eqnarray}
where dots denote derivatives with respect to physical time, $t$, $H = \dot{a}/a$ is the Hubble parameter, and $f_{\mathcal{R}}$ is the extra scalar degree of freedom introduced in the model. Here, $f_{\mathcal{R}}$, $f_{\mathcal{RR}}$, $f_{\mathcal{RRR}}$ denote the first, second and third derivatives of $f(\mathcal{R})$ with respect to $\mathcal{R}$.
Eqs.~(\ref{hyb1}), (\ref{hyb2}) and (\ref{hyb3}) constitute a closed set of differential equations that determines the background evolution for specified $f(\mathcal{R})$.
Note that we recover the standard Friedmann equations of $\Lambda$CDM in the limit of $f_{\mathcal{R}} \rightarrow 0$.

Lastly, it is useful to write the effective mass of the additional scalar degree of freedom, which is given by~\cite{main1,main2}
\begin{equation}
m^{2}_{f_{\mathcal{R}}} = \frac{2V(f_{\mathcal{R}})- V_{f_{\mathcal{R}}}-f_{\mathcal{R}}\PC{1+f_{\mathcal{R}}}V_{f_{\mathcal{R}}f_{\mathcal{R}}}}{3},
\end{equation}
where $V(f_{\mathcal{R}}) = \mathcal{R}f_{\mathcal{R}} - f(\mathcal{R})$ is the scalar field potential, defined in the scalar-tensor formulation of the hybrid metric-Palatini theory.
%
%\vanessa{Think about how to add the motivations of the introduction in here. Do not over mention arguments.}

The hybrid metric-Palatini theory avoids the well known instabilities in the pure Palatini approach \cite{Capozziello:2015lza,palainst1,palainst2} by providing a propagating additional scalar degree of freedom that can modify gravity across all scales due to its light, long-range interacting nature \cite{main1,main2,hybridcosmo,hybridgala}. Furthermore, it does not require the effective mass of the scalar field to be massive in order to be viable on small scales, as this is assured as long as the amplitude of the scalar field remains small \cite{main2}, which our designer model naturally guarantees.

\subsubsection*{Designer $f(\mathcal{R})$ Model} \label{model1}

We briefly review the designer hybrid-metric Palatini model that we will adopt to describe the evolution of the decaying early modification of gravity and its observational constraints in Sec.~\ref{II}. This model was first introduced in Ref.~\cite{hybridpert}, and it allows one to retrieve a family of $f(\mathcal{R})$ functions that produce a background evolution indistinguishable to $\Lambda$CDM from solving the second-order differential equation
\begin{equation}{\label{Fdiff}}
 f_{\mathcal{R}}^{\prime \prime} + f_{\mathcal{R}}^{\prime}\PC{\frac{E^{\prime}}{2E} - 1} + f_{\mathcal{R}} \frac{E^{\prime}}{E} - \frac{3}{2}\frac{f_{\mathcal{R}}^{\prime 2}}{f_{\mathcal{R}}} = 0 \,,
\end{equation}
where here and throughout the rest of the paper primes represent derivatives with respect to $\ln a$. Eq.~(\ref{Fdiff}) is obtained from setting the effective equation of state $w_{\rm{eff}}$ equal to $-1$. The background evolution is fixed through $E\PC{a} \equiv H^2/H_{0}^{2} = \Omega_{\rm{m}}a^{-3} + \Omega_{\rm r}a^{-4} + \Omega_{\textrm{eff}}a^{3 \int_{a}^{1} \PC{1+w_\textrm{eff}}d\ln a}$.
In a flat Universe, $\Omega_{\textrm{eff}} = 1 - \Omega_{\rm{m}} - \Omega_{\rm r}$ and, for $w_{\textrm{eff}} = -1$, one recovers a $\Lambda$CDM-like background cosmology.
The initial conditions for solving Eq.~(\ref{Fdiff}) are set at an initial scale factor, $a_{\rm i}=(1+z_{\rm i})^{-1} \ll 1$, by~\cite{hybridpert}
\begin{eqnarray}
 f_{\mathcal{R}\rm{i}} & = & C_{1} a_{\rm i}^{-a_{\rm{aux}}} \PR{\cosh \PC{\frac{1}{2}\PR{ \ln a_{\rm i} + C_{2}} \sqrt{d}}}^{-2} \,, \label{initialF} \\
 f_{\mathcal{R}\rm{i}}^{\prime} & = & - C_{1} \frac{a_{\rm i}^{-a_{\rm{aux}}}}{\cosh \PC{...}^{2}} \PR{a_{\rm{aux}} + \sqrt{d} \tanh \PC{...}} \,, \label{initialFprime}
\end{eqnarray}
where $d = a_{\rm{aux}}^{2} - 2b$,  $a_{\rm{aux}} = \PC{5+6r_{\rm i}}/\PC{2 + 2r_{\rm i}}$ and $b = \PC{3+4r_{\rm i}}/\PC{1+r_{\rm i}}$, with $r_{\rm i} = \Omega_{\gamma} \PC{\Omega_{\rm{m}} a_{\rm i}}^{-1}$. The dotted arguments of the hyperbolic functions refer to the same argument as in the hyperbolic cosine in Eq.~(\ref{initialF}).
Throughout the paper we fix $C_{\rm 2}$ to a large value in order for the absolute value of the hyperbolic tangent to be close to unity. $C_{\rm 1}$ is then fixed by choosing a value for $f_{\mathcal{R}\rm{i}} \equiv f_{\mathcal{R}}(z=z_{\rm{i}})$. Hence, one then just has to numerically evolve the model using Eq.~(\ref{Fdiff}), and make use of the background equations to recover further quantities of interest, such as $f(\mathcal{R})$, at each step of the iteration.

\begin{figure}[t!]
\begin{center}
\includegraphics[scale = 0.47]{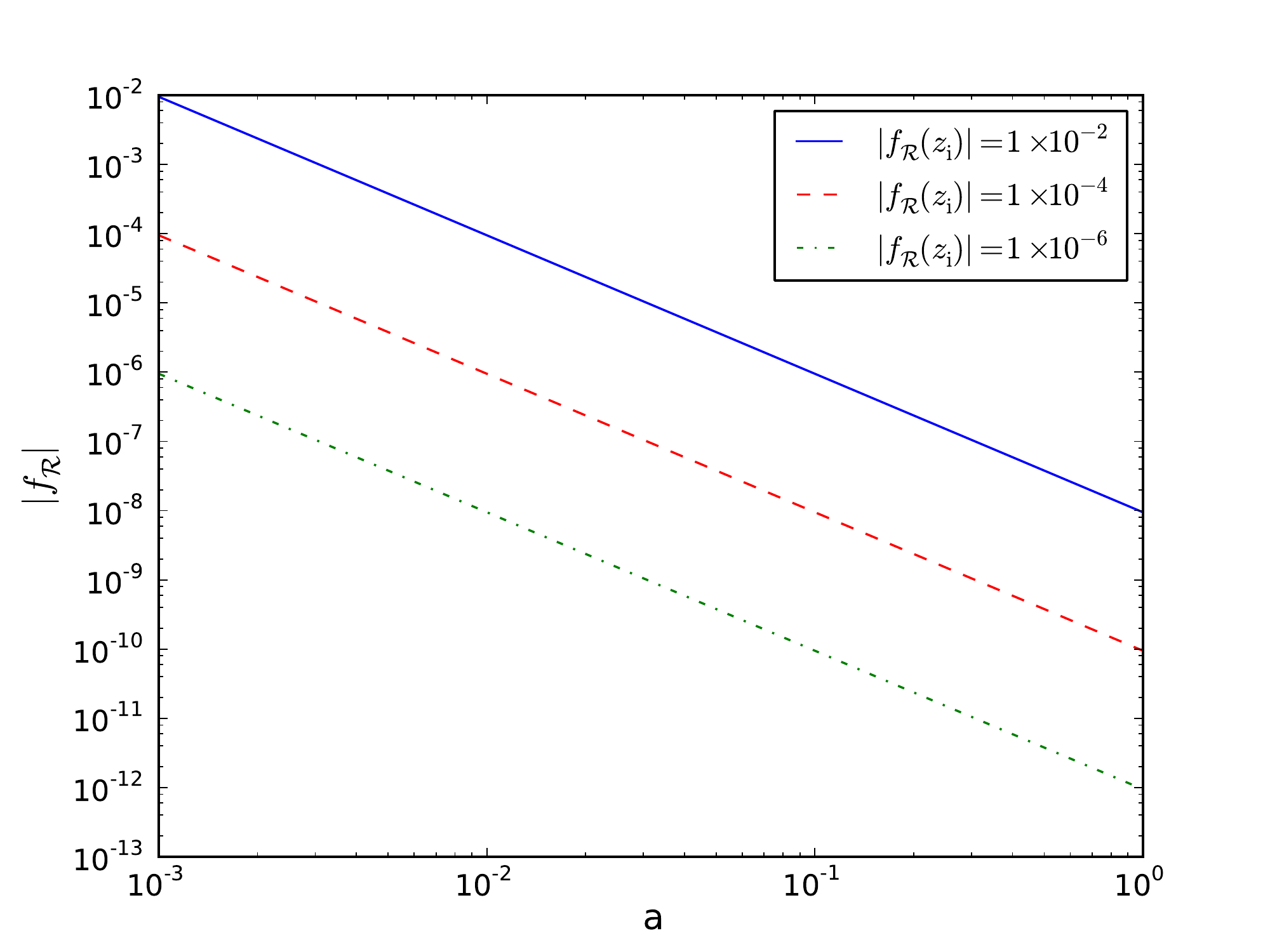}
\end{center}
\caption{\label{f_r_compare} Evolution of the absolute value of the extra scalar degree of freedom introduced in $f(\mathcal{R})$ theories, $f_{\mathcal{R}}$, as a function of the scale factor, $a$, with $z_{\rm{i}}=1000$. We have fixed $\Omega_{\rm{m}}=0.30$ for illustration.
}
\end{figure}

In Fig.~\ref{f_r_compare} we plot the evolution of the absolute value of $f_{\mathcal{R}}$ as a function of the scale factor for different initial values $f_{\mathcal{R}\rm{i}}$ set at a redshift $z_{\rm{i}} = 1000$.
The scalar field $f_{\mathcal{R}}$ decays with time and is strongly suppressed as we approach $a \rightarrow 1$.
In Sec.~\ref{II} it will become evident that due to this suppression, $f(\mathcal{R})$ behaves like a decaying early-modified gravity model that satisfies Solar-System constraints~\cite{main1}.

Having a hybrid metric-Palatini model that recovers a $\Lambda$CDM-like background evolution allows to separate the modifications introduced between linear perturbations from background effects. Possible deviations at the background level from $\Lambda$CDM for other $f(\mathcal{R})$ functions have already been tested against observations in Ref.~\cite{hybridback}, where constraints on the initial value of the scalar field $f_{\mathcal{R}i}$ at very early times ($z_{\rm i}\sim10^8$) were obtained, restricting its maximum amplitude to an absolute value of $1 \times 10^{-2}$. Considering how hybrid metric-Palatini models couple to the gravitational sector through a factor of $\PC{1+\phi}$, this constrained the maximum variation of the effective gravitational constant $G_{\rm{eff}}$ in the background evolution to $1 \%$ of its Newtonian value. This result is compatible with constraints on $G_{\rm{eff}}$ from big bang nucleosynthesis \cite{mgc1}.

Modifications introduced in the linear cosmological perturbations have not yet been tested for $f(\mathcal{R})$ gravity, and the designer model discussed here perfectly suits this purpose. As we will show in Sec.~\ref{constraints}, the constraints we obtain in this work using CMB temperature and polarization anisotropy data are in perfect agreement with existent (and expected) constraints on $G_{\rm{eff}}$ considering the same effects \cite{mgc2}.

\subsection{Linear Perturbations in $f(\mathcal{R})$ Gravity} \label{perturbations}

We briefly review the main aspects concerning the evolution of linear perturbations in the hybrid metric-Palatini theory. For the full set of linearly perturbed Einstein and scalar field equations we direct the reader to Ref.~\cite{hybridpert}. 
Typically for modified gravity theories (however, see Refs.~\cite{lombriser:14b,lombriser:15b}), the hybrid metric-Palatini theory introduces a non-zero slip between the gravitational potentials in the Newtonian gauge, $\Phi=\delta g_{00}/(2g_{00})$ and $\Psi=-\delta g_{\rm ii}/(2g_{\rm ii})$.
Neglecting any anisotropic contribution from matter fields, the anisotropy equation becomes
\begin{equation}{\label{anisotropic}}
 \Phi - \Psi = \frac{\delta f_{\mathcal{R}}}{1+f_{\mathcal{R}}} \,,
\end{equation}
where $\delta f_{\mathcal{R}}$ is the linear perturbation of the scalar field with its background value denoted by $f_{\mathcal{R}}$. 
The evolution of $\delta f_{\mathcal{R}}$ is dictated by the linear perturbation of the scalar field equation of motion,
\begin{eqnarray}{\label{scalarnewton}}
\ddot{\delta f_{\mathcal{R}}} + \dot{\delta f_{\mathcal{R}}} \PC{2 \mathcal{H} - \frac{\dot{f_{\mathcal{R}}}}{f_{\mathcal{R}}}} + & & \nonumber\\
 \delta f_{\mathcal{R}} \PC{k^2 + \frac{\dot{f_{\mathcal{R}}}^2}{2 {f_{\mathcal{R}}}^2} + a^2 m_{f_{\mathcal{R}}}^{2} - \frac{\kappa^2}{3} a^2 T } + & & \nonumber\\
 \Psi \PC{ \frac{\dot{f_{\mathcal{R}}}^2}{f_{\mathcal{R}}} - 2 \ddot{f_{\mathcal{R}}} - 4 \dot{f_{\mathcal{R}}} \mathcal{H} } - \dot{f_{\mathcal{R}}}\PC{3\dot{\Phi} + \dot{\Psi}} & = & \frac{f_{\mathcal{R}}}{3}a^2 \kappa^2 \delta T , \nonumber\\
\end{eqnarray}
where $\delta T$ denotes the linear perturbation of the trace of the stress-energy tensor, $T = -\rho + 3p$, and {\it{for this equation only}}, dots indicate derivatives with respect to conformal time $\tau$ with $dt = a \: d\tau$, and $\mathcal{H} \equiv a H$.

It has been shown in Ref.~\cite{hybridpert} that the evolution of $\delta f_{\mathcal{R}}$ is characterized by quick oscillations around zero, which end up reflecting in the ratio between the Newtonian potentials, $\gamma\equiv\Phi/\Psi$. These oscillations are scale dependent, oscillating faster and  with larger amplitude at smaller scales. They can produce noticeable oscillations at near-horizon scales, depending on the initial value of the scalar field at early times that, for instance, have an impact on the Poisson equation. Due to the Hubble friction term (see Eq.~(\ref{scalarnewton})), these modifications eventually get damped as one approaches $a \approx 1$, becoming negligible at the present with no signs of significant subhorizon modifications. 

We will explore the behavior of $\delta f_{\mathcal{R}}$ further in Secs.~\ref{subapp} and \ref{earlyapp}, focusing on its subhorizon and early-time evolution, respectively, where we will develop accurate approximations for these regimes. In order to test our approximations, we follow Ref.~\cite{hybridpert} and solve the exact numerical evolution of the gravitational potentials and $\delta f_{\mathcal{R}}$, using the linearly perturbed conservation equations for the stress-energy tensor and the first-order differential equations for the lensing potential, $\Phi_{+} \equiv \PC{\Phi + \Psi}/2$.

\subsubsection{Subhorizon Approximation} \label{subapp}

\begin{figure*}[t!]
$
\begin{array}{cc}
\includegraphics[scale = 0.4]{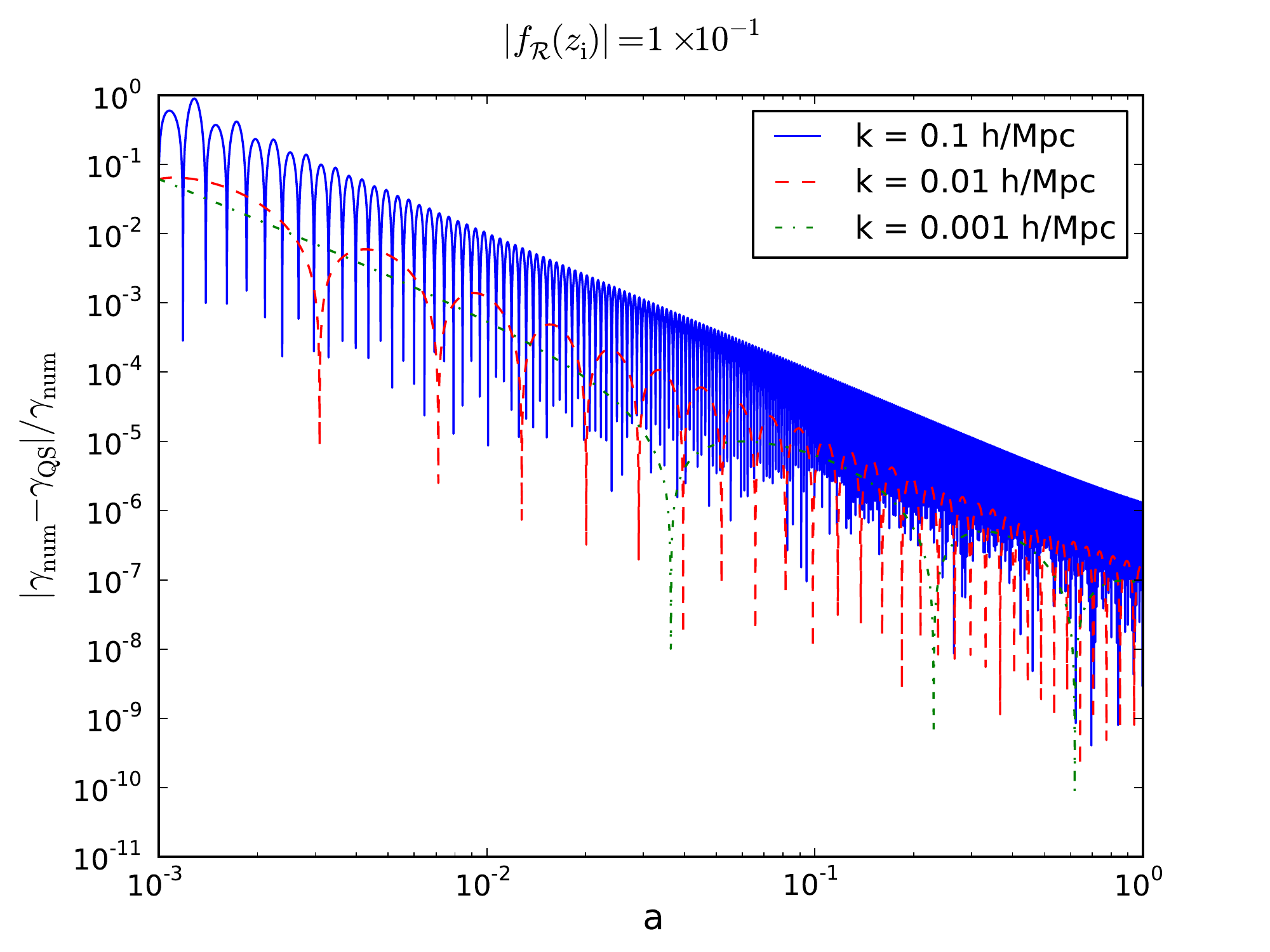} & \includegraphics[scale=0.4]{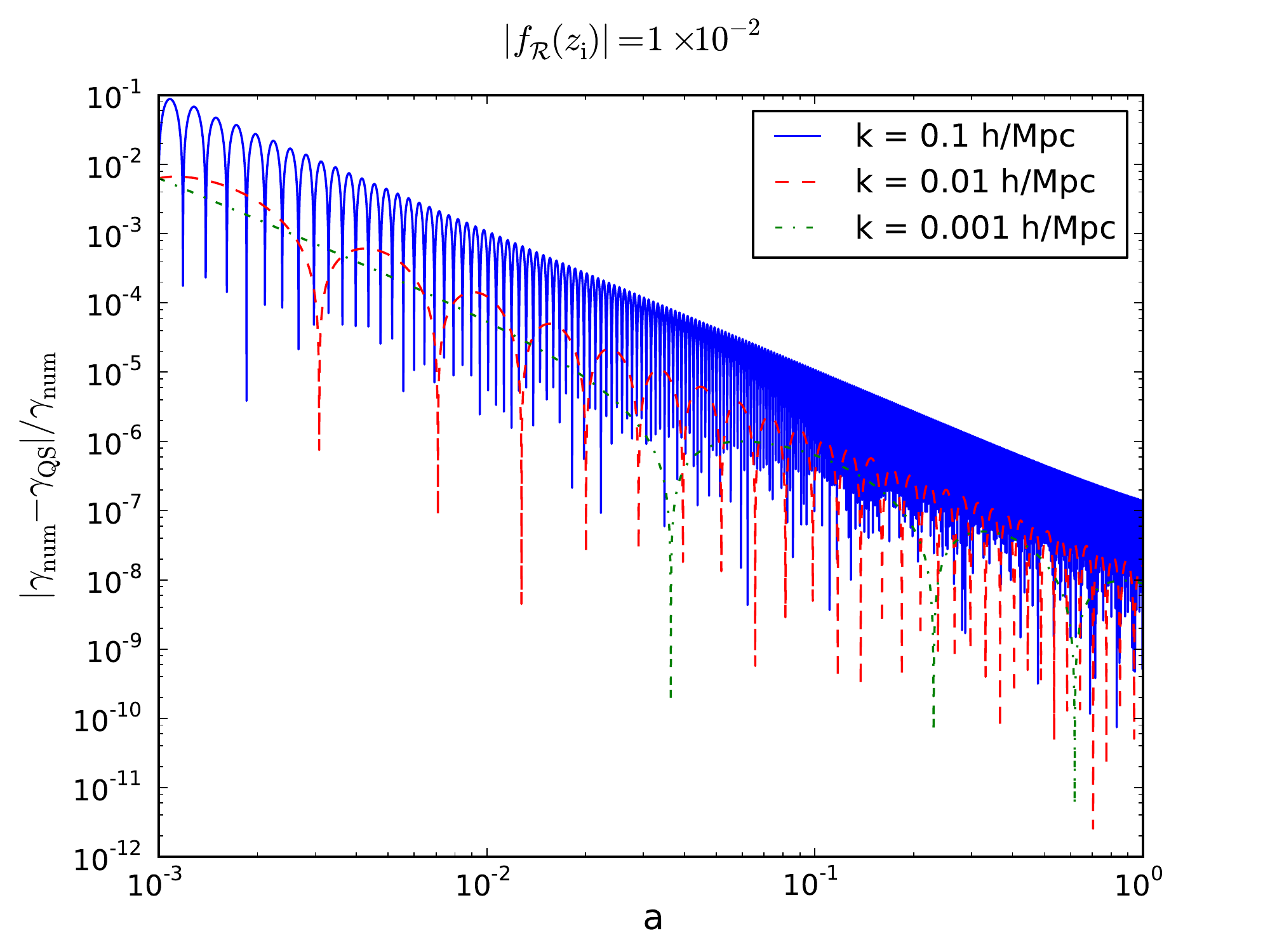}\\
\includegraphics[scale = 0.4]{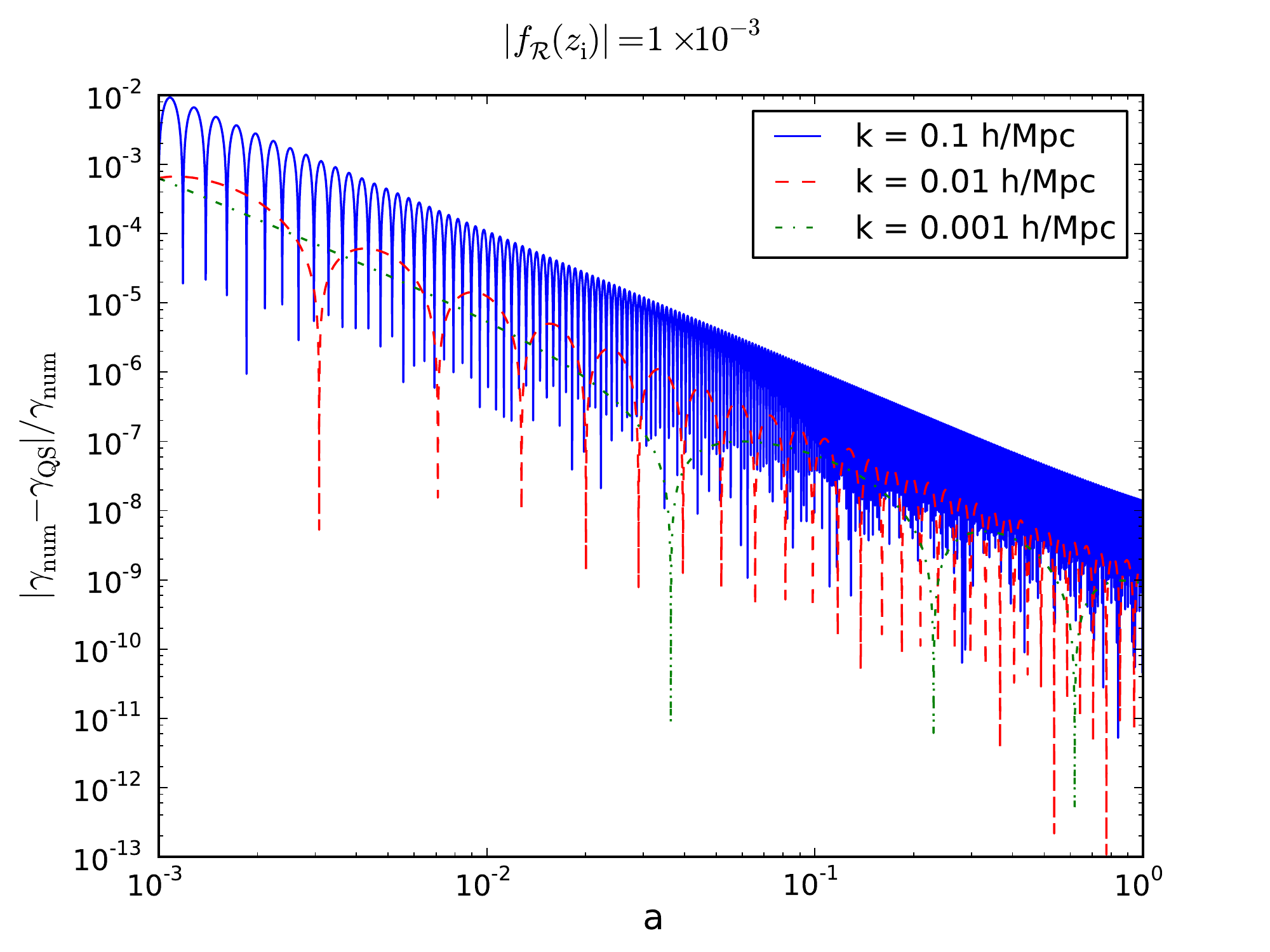} & \includegraphics[scale=0.4]{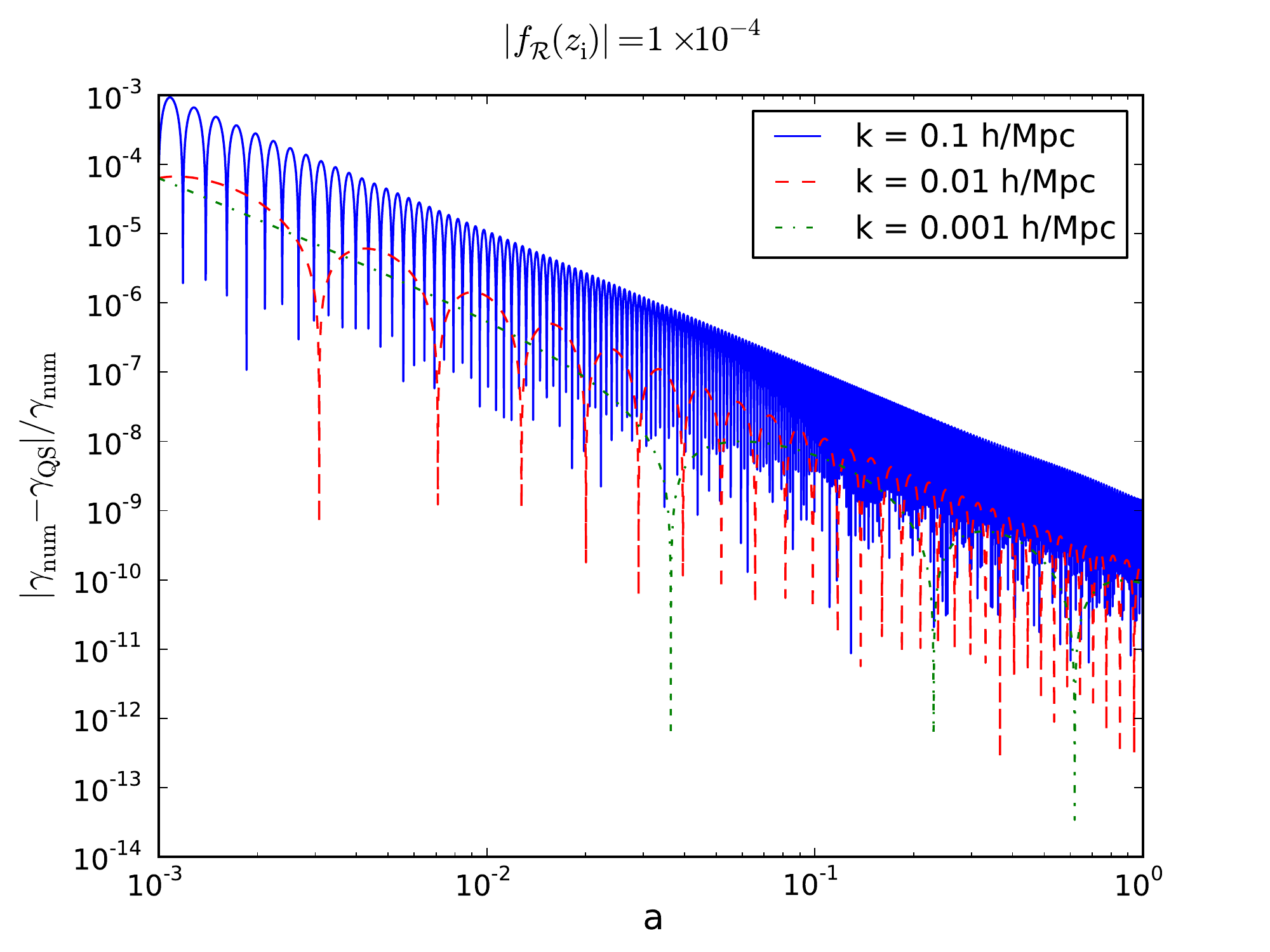}
\end{array}$
\caption{\label{figureqs}
Relative difference $|\gamma_{\rm{num}} - \gamma_{\rm{QS}}|/\gamma_{\rm{num}}$ between the numerical ratio $\gamma\equiv\Phi/\Psi$ and its quasistatic (QS) approximation given by Eq.~(\ref{qsratio}). We have considered $z_{\rm{i}} = 1000$ and fixed $\Omega_{\rm{m}} = 0.30$.}
\end{figure*}

We first consider wavemodes that are deep within the Hubble radius with wavenumber $k \gg aH$. To describe this limit, we adopt the quasistatic approximation, discarding time derivatives of perturbations when compared to their spatial variation. Generally, for Horndeski scalar-tensor theories, this is a good approximation on small scales~\cite{lombriser:15a}.
In practice, this allows one to keep the terms proportional to $k^2/\PC{a^2 H^2}$ as well as those related to the matter perturbation $\delta \rho_{\rm{m}}$ and the scalar field effective mass $m_{f_{\mathcal{R}}}^{2}$. The latter sets a modified length scale that can be compared to that of the perturbations.

From the $0-0$ linearly perturbed Einstein equation in the Newtonian gauge, we obtain in the subhorizon regime~\cite{hybridpert}
\begin{equation}{\label{phiqs1}}
 \frac{k^2}{a^2} \Phi \approx \frac{1}{2\PC{1+f_{\mathcal{R}}}}\PR{\delta f_{\mathcal{R}} \PC{\frac{k^2}{a^2}} - \kappa^2 \delta \rho_{\rm{m}}}\,,
\end{equation}
where $\delta \rho_{\rm{m}} \equiv \rho_{\rm{m}} \delta_{\rm{m}}$. Using this approximation in the anisotropy equation we then get
\begin{equation}{\label{psiqs1}}
\frac{k^2}{a^2} \Psi \approx -\frac{1}{2\PC{1+f_{\mathcal{R}}}}\PR{\delta f_{\mathcal{R}} \PC{\frac{k^2}{a^2}} + \kappa^2 \delta \rho_{\rm{m}}}\,.
\end{equation}
One can then calculate a similar approximation for $\delta f_{\mathcal{R}}$ from Eq.~(\ref{scalarnewton}),
\begin{equation}{\label{delphiqs}}
 \delta f_{\mathcal{R}} \approx - \frac{H_0^{2}E_{\rm{m}}}{k^2/a^2 + m_{f_{\mathcal{R}}}^{2}} f_{\mathcal{R}} \hspace{0.3 mm} \delta_{\rm{m}} \,,
\end{equation}
which can be inserted back into Eqs.~(\ref{phiqs1}) and (\ref{psiqs1}) such that
\begin{eqnarray}
\frac{k^2}{a^2} \Phi &\approx& - \frac{H_0^{2}E_{\rm{m}}\delta_{\rm{m}}}{2\PC{1+f_{\mathcal{R}}}}\PR{\frac{k^2/a^2\PC{f_{\mathcal{R}} + 3} + 3m_{f_{\mathcal{R}}}^{2}}{k^2/a^2 + m_{f_{\mathcal{R}}}^{2}}},{\label{phiqs2}} \\
\frac{k^2}{a^2} \Psi &\approx& - \frac{H_0^{2}E_{\rm{m}}\delta_{\rm{m}}}{2\PC{1+f_{\mathcal{R}}}}\PR{\frac{k^2/a^2\PC{3 - f_{\mathcal{R}}} + 3m_{f_{\mathcal{R}}}^{2}}{k^2/a^2 + m_{f_{\mathcal{R}}}^{2}}},{\label{psiqs2}}
\end{eqnarray}
where $E_{\rm{m}} \equiv \Omega_{\rm{m}} a^{-3}$.

These approximations can, in turn, be used to obtain an expression for the lensing potential, $\Phi_{+}$, in this regime:
\begin{equation}{\label{lenqs}}
 \frac{k^2}{a^2} \Phi_{+} \approx - \frac{3H^{2}_0 E_{\rm{m}}}{2\PC{1+f_{\mathcal{R}}}} \delta_{\rm{m}}
\end{equation}
whereas the slip between the potentials, $\delta f_{\mathcal{R}}$, is given by
\begin{equation}{\label{chiqs}}
 \delta f_{\mathcal{R}} \approx \frac{2}{3} \frac{k^2}{a^2}\frac{f_{\mathcal{R}} \hspace{0.3 mm} \Phi_{+}}{k^2/a^2 + m_{f_{\mathcal{R}}}^{2}} \,.
\end{equation}

As mentioned in Sec.~\ref{Int}, the background value of the scalar field is required to be small in order for the metric-Palatini theory to avoid Solar-System tests. In these circumstances, the quasistatic modifications will be almost unnoticeable, even if the range of the modifications, given by the effective Compton wavelength $ \lambda_{\rm C} = 2 \pi / m_{f_{\mathcal{R}}}$, is relevant. For instance, note that for $f_{\mathcal{R}} \rightarrow 0$, $\delta f_{\mathcal{R}} \rightarrow 0$ since $\delta f_{\mathcal{R}}$ is proportional to the background value of the scalar field $f_{\mathcal{R}}$ in the quasistatic regime, as can be seen in Eq.~(\ref{chiqs}).

The $f(\mathcal{R})$ models that have been analyzed so far~\cite{hybridback,hybridpert} evolve towards smaller deviations from $\Lambda$CDM as we approach the present, with $f_{\mathcal{R}}$ tending to negligible values. This renders the modifications in the quasistatic regime subdominant, as was explicitly shown in Ref.~\cite{hybridpert} for the designer $f(\mathcal{R})$ model, with no mentionable enhancement of the perturbations in this regime when compared to $\Lambda$CDM.

In Fig.~\ref{figureqs} we compare the numerical evolution of the ratio between the Newtonian potentials, $\gamma$, with its quasistatic approximation,
\begin{equation}{\label{qsratio}}
 \gamma_{\rm{QS}} \equiv \frac{\Phi}{\Psi} = \frac{k^2/a^2\PC{3+f_{\mathcal{R}}}+3m_{f_{\mathcal{R}}}^{2}}{k^2/a^2\PC{3-f_{\mathcal{R}}}+3m_{f_{\mathcal{R}}}^{2}} \,.
\end{equation}
We see that it is an accurate approximation at late times, as a consequence of large $k/(aH)$ values. As we approach the present time in our models, the subhorizon modifications become suppressed, leading in turn to a very small difference between the compared values. This accuracy holds even when we consider larger initial values for the scalar field, $f_{\mathcal{R}{\rm i}}$.

%However, the quasistatic approximation breaks down at earlier times, due to the oscillatory behavior of $\delta f_{\mathcal{R}}$ discussed in Sec.~\ref{perturbations}. This becomes more evident at smaller scales, where the amplitude of the oscillations is larger.
%\lucas{If the effect goes as $k/a$, isn't small scales and early times somewhat interchangeable?}

%\vanessa{However, the quasistatic approximation breaks down at earlier times, due to the oscillatory behavior of $\delta f_{\mathcal{R}}$ discussed in Sec.~\ref{perturbations}. This is a result from the $k/(aH)$ behavior, causing the amplitude of the oscillations to be larger. }
However, the quasistatic approximation breaks down at earlier times for the smaller scales, due to the oscillatory behavior of $\delta f_{\mathcal{R}}$ discussed in Sec.~\ref{perturbations}. For large initial values of the scalar field the error can be of order unity and decreases as we consider smaller values for $f_{\mathcal{R}i}$.
Hence, for an accurate but computationally efficient description of the evolution of $\gamma$ in the designer $f(\mathcal{R})$ model that is valid across a large range of redshifts and scales, some corrections must be applied to the subhorizon approximation (see Sec.~\ref{constraints}).

Lastly, we emphasize that in the hybrid metric-Palatini model, $f_{\mathcal{R}}$ and $\delta f_{\mathcal{R}}$ are strongly suppressed at the present, and $(G_{\rm{eff}}-G)/G\ll 1$ at any scale, consistent with Solar-System tests.
In contrast, in metric $f(R)$ gravity, for modes well within the Compton radius, we have $(G_{\rm{eff}}-G)/G=4/3$ at linear order, and the model needs to employ a nonlinear chameleon mechanism~\cite{khoury:03,hu:07,brax:08,lombriser:14a} to restore $G_{\rm{eff}}/G \rightarrow 1$ at the small scales probed by Solar-System tests.
%\lucas{Now we mention something similar at the beginning of the section.}
%
%
%However, in chameleon models, for a given scalar field value, there always exist small objects in low-density environments at late times below a characteristic threshold that remain unscreened and, hence, are affected by the gravitational modification.
%These are potentially observable and could be used to constrain the modification.
%In contrast, these objects do not experience an enhanced $G_{\rm eff}$ in the hybrid metric-Palatini scenario and thus even if observable would not 
%
Unlike the chamelon mechanism, however, the suppression in the hybrid metric-Palatini model is independent of environment and cannot be tested by unscreened small objects in voids~\cite{dwarf_galaxies} or unscreened by environment-dependent statistical measurements of the large-scale structure~\cite{lombriser:15}.
We recall that it is for this aspect that we adopt the decaying early-time gravitational modification characterized by the hybrid metric-Palatini model rather than the decaying mode of metric $f(R)$ gravity, where an effective 4/3 enhancement of the gravitational coupling at late times would always be present at some level.
%
%Note that the suppression is also independent of environment and cannot be unscreened by environment-dependent statistical measurements of the large-scale structure~\cite{lombriser:15} or tested by unscreened small objects in voids~\cite{dwarf_galaxies}.

\subsubsection{Early--Time Corrections} \label{earlyapp}

\begin{figure*}[t!]
$
\begin{array}{cc}
\includegraphics[scale = 0.45]{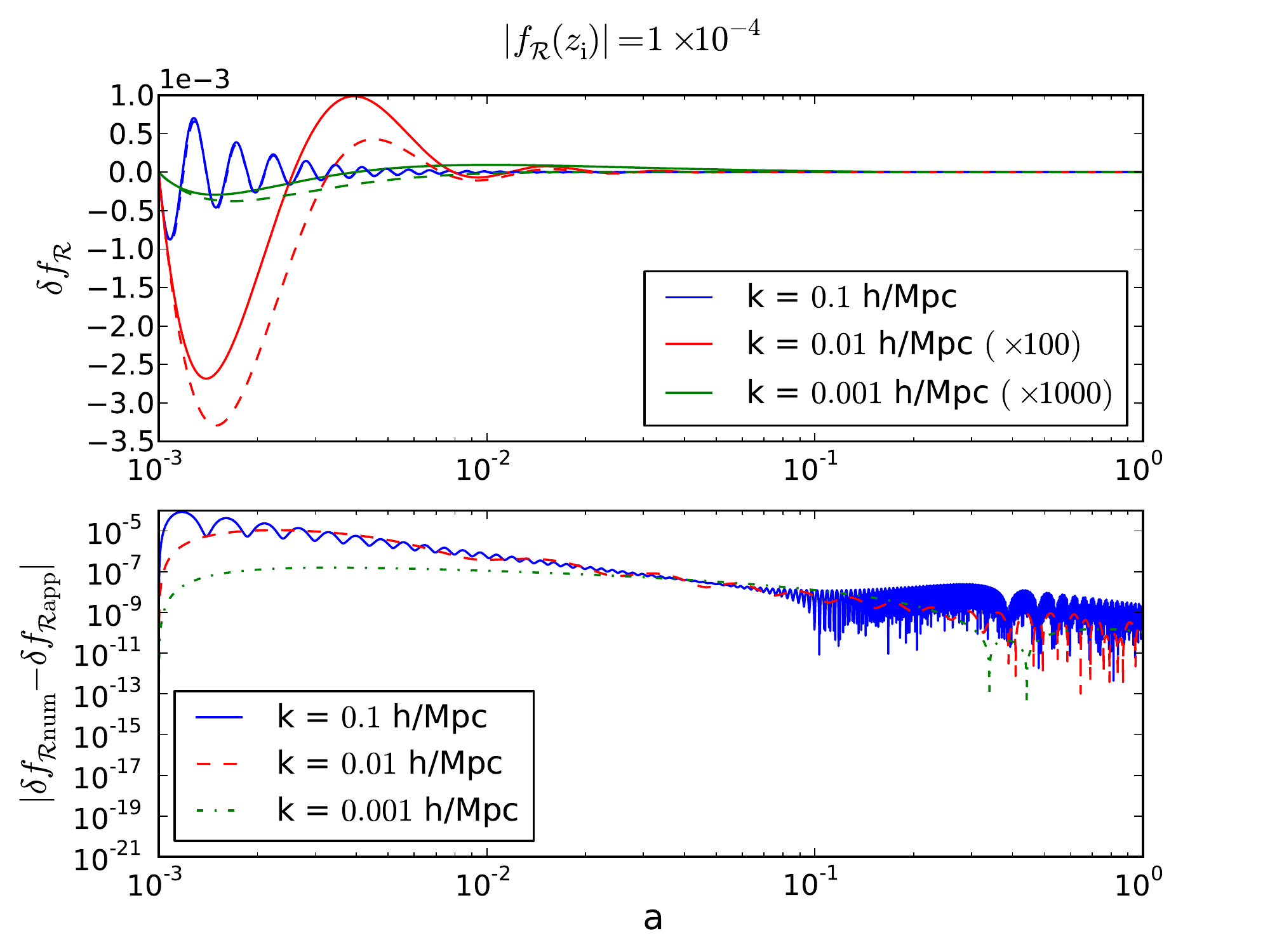} & \includegraphics[scale = 0.45]{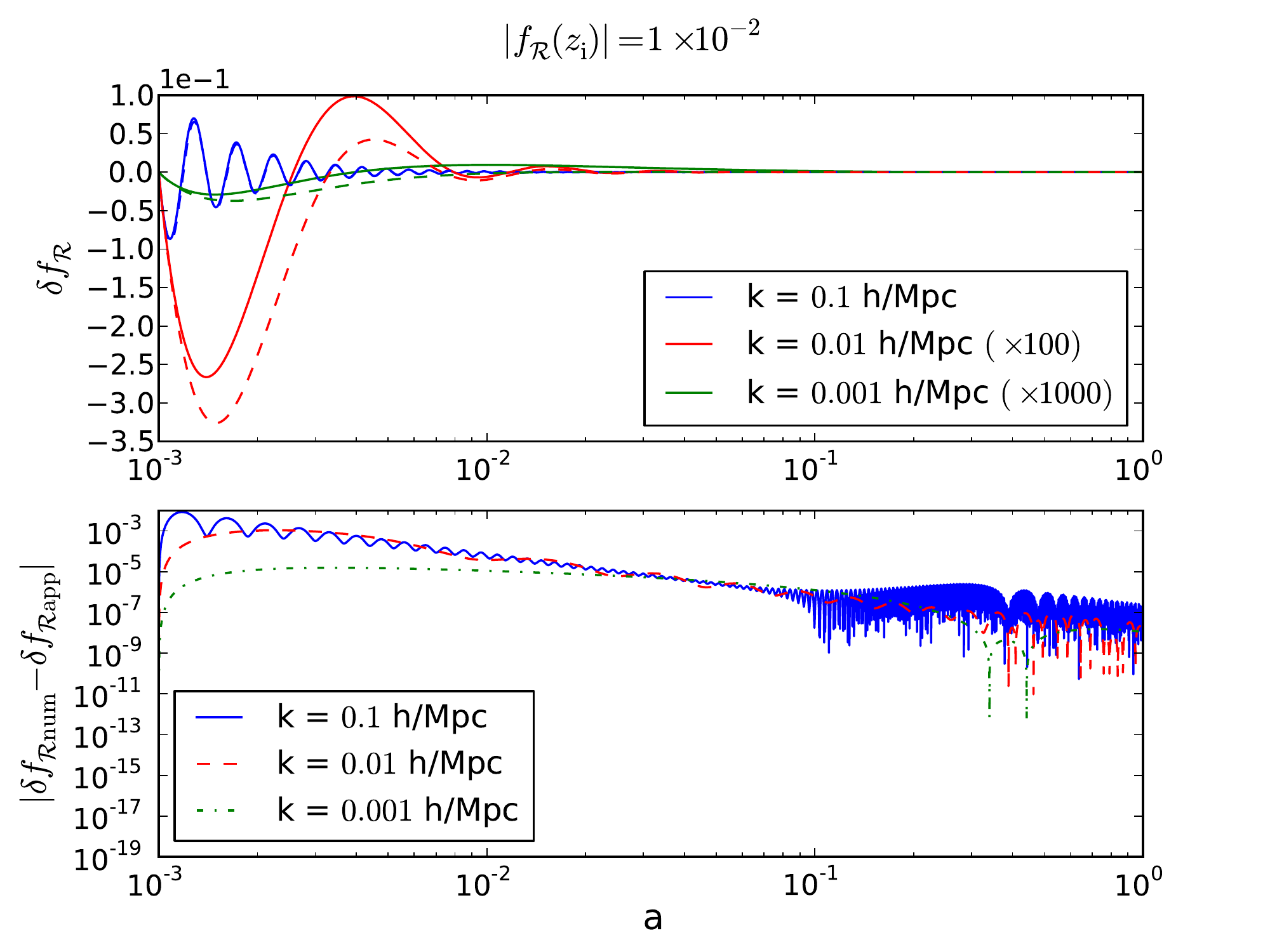}
\end{array}$
\caption{\label{delphicomp}The top panels show the numerical evolution (solid lines) of the perturbation $\delta f_{\mathcal{R}}$ against the evolution predicted by our analytical approximation (dashed lines) given by Eq.~(\ref{delphiwkbapp}). The two largest scales have been enhanced by a factor of $100$ and $1000$ to be noticeable. The bottom panels show the absolute difference between the analytical approximation and the numerical results. We have fixed $\Omega_{\rm{m}} = 0.30$.}
\end{figure*}

The dynamics of $\delta f_{\mathcal{R}}$ is dictated by Eq.~(\ref{scalarnewton}), which is the equation of a damped harmonic oscillator with a driving force proportional to the matter perturbation.
The frequency of the oscillation
depends on the mode wavenumber $k$, while the damping term is dominated by the Hubble parameter at early times, and $\delta f_{\mathcal{R}}$ quickly becomes negligible towards late times, where the oscillations are no longer observable. The driving term could deviate the equilibrium position of the oscillations. However, note that it is proportional to $f_{\mathcal{R}}$, which not only is fixed to a small value at early times as we study small deviations from GR, but also evolves towards zero at late times, rendering the external force term almost negligible.

Hence, rewriting Eq.~(\ref{scalarnewton}) to depend on $\ln a$, assuming ${f_{\mathcal{R}},\hspace{0.25 mm} \dot{f_{\mathcal{R}}}} \ll 1$, but not neglecting terms proportional to $\dot{f_{\mathcal{R}}}/f_{\mathcal{R}}$, we approximate it to
\begin{eqnarray}{\label{scalarnewtonapp}}
 \delta f_{\mathcal{R}}^{\prime \prime} &&+ \delta f_{\mathcal{R}} \PC{\frac{k^2}{a^2H^2} + \frac{f_{\mathcal{R}}^{\prime 2}}{2 f_{\mathcal{R}}^{2}} + \frac{H^{2}_{0}\Omega_{\rm{m}}a^{-3}}{H^{2}}} + \nonumber \\
 &&+ \delta f_{\mathcal{R}}^{\prime} \PC{3 + \frac{H^{\prime}}{H} - \frac{f_{\mathcal{R}}^{\prime}}{f_{\mathcal{R}}}} \approx 0 \,,
\end{eqnarray}
for which we attempt a solution under the Wentzel--Kramers--Brillouin (WKB) approximation given by
\begin{equation}{\label{delphiwkbapp}}
 \delta f_{\mathcal{R}} \approx \frac{A}{\sqrt{2w}} a^{-\gamma_{\rm{exp}}} \cos \PC{\int w \hspace{0.3 mm} d\ln a + \theta_0}.
\end{equation}
We expect the approximation to be valid as long as the adiabatic condition $|\dot{w}| \ll w^2$ holds, where $w^2$ is the term multiplying $\delta f_{\mathcal{R}}$ in Eq.~(\ref{scalarnewtonapp}); and $\gamma_{\rm{exp}}$ is the quantity multiplying the $\delta f_{\mathcal{R}}^{\prime}$ term in Eq.~(\ref{scalarnewtonapp}). The constants $\theta_0$ and $A$ can be fixed by imposing suitable initial conditions for $\delta f_{\mathcal{R}}$ and $\delta f_{\mathcal{R}}^{\prime}$ at a chosen redshift.

For the $f(\mathcal{R})$ designer model, the ratio between $f_{\mathcal{R}}^{\prime}$ and $f_{\mathcal{R}}$ can be easily calculated at early times using the initial conditions presented in Sec.~\ref{I}.
This yields
\begin{equation}{\label{ratiophipphi}}
 \frac{f_{\mathcal{R}}^{\prime}}{f_{\mathcal{R}}} \approx \sqrt{d} - a_{\rm{aux}} \,,
\end{equation}
With this approximation, it is possible to simplify $w$ and obtain an analytical solution for the integral $\int w\hspace{0.3 mm} d  \ln a$. The details of this calculation may be found in Appendix~\ref{appendix1}.

In Fig.~\ref{delphicomp} we set the initial conditions for $\delta f_{\mathcal{R}}$ by determining $\theta_0$ such that $\delta f_{\mathcal{R}}$ is zero at the chosen initial redshift $z_{\rm i} = 1000$. We note that this is completely arbitrary, but not particularly relevant for the overall evolution of $\delta f_{\mathcal{R}}$ since it quickly oscillates around zero. We can then differentiate Eq.~(\ref{scalarnewtonapp}) with respect to $\ln a$ and compute $A$ by calculating the numerical value of $\delta f_{\mathcal{R}}^{\prime}$ using Eq.~(67) of Ref.~\cite{hybridpert} at the same redshift.

We see in Fig.~\ref{delphicomp} that our analytical approximation works remarkably well, considering the complexity of the equation describing the dynamics of $\delta f_{\mathcal{R}}$. Even though it may fail in predicting the exact amplitude of the oscillations, the relative difference to the numerical results is insignificantly small compared to the precision available with current experiments. Also, it clearly encompasses the desired dependence on the scale of the modes of the perturbations, with a higher amplitude and frequency of oscillation the smaller scales (higher $k$) one considers.

Lastly, Fig.~\ref{delphicomp} serves as further confirmation of the viability of the subhorizon approximations derived in Sec.~\ref{subapp} at late times.
As Eq.~(\ref{chiqs}) dictates, $\delta f_{\mathcal{R}}$ should be strongly suppressed in the subhorizon regime following the behavior of the background scalar field value and with $k \gg aH$.

\subsection{Decoupling at High Redshifts} \label{sec:decoupling}

The hybrid metric-Palatini modification of gravity needs to decouple at high redshifts in order not to violate stringent high-curvature constraints from the CMB.
However, we wish to determine below which redshift $z_{\rm on}$ the modification can be introduced and to which degree a decaying early-time modification motivated by the evolution of hybrid metric-Palatini gravity at $z\leq z_{\rm on}$ can be constrained by the CMB radiation observed today.
In order to formulate an explicit realization of the decaying early modified gravity model, we embed the designer hybrid metric-Palatini scenario with high-redshift decoupling in Horndeski scalar-tensor theory~\cite{horndeski:74} using the effective field theory of cosmic acceleration (see Ref.~\cite{gleyzes:14} for a review).

\subsection{Embedding in Horndeski Gravity and Effective Field Theory} \label{coneft}

We will now embed the designer $f(\mathcal{R})$ model in the effective field theory of Horndeski gravity following the notation of Ref.~\cite{eftformalism}.
%We now proceed to outline how the designer hybrid metric-Palatini model, detailed in Sec.~\ref{I}, can be embedded in the Horndeski scalar-tensor theory.
%We use the effective field theory of cosmic acceleration, where we adopt the notation of Ref.~\cite{eftformalism}.
%\lucas{Thes sentences are very repetitive. The last sentence of II.C is saying almost the same. Try shorten it while keeping the references to II.A and 43.}
Given the $\Lambda$CDM background expansion history of our designer hybrid metric-Palatini model, its modifications are fully specified by the effective parameters characterizing the linear perturbations,
\begin{equation}
 \alpha_{\rm M} = \frac{f_{\mathcal{R}}'}{1+f_{\mathcal{R}}}, \ \ \ \
 \alpha_{\rm K} = -\frac{3}{2}\frac{f_{\mathcal{R}}'}{f_{\mathcal{R}}}\alpha_{\rm M}, \ \ \ \
 \alpha_{\rm B} = -\alpha_{\rm M}, \label{eq:eft}
\end{equation}
where $\alpha_{\rm M}\equiv(M_{*}^{2})'/M_{*}^{2}$ describes the running of the Planck mass $\kappa^2 M_{*}^{2} \equiv 1 + f_{\mathcal{R}}$; $\alpha_{\rm K}$ denotes the contribution of the kinetic energy of the scalar field; and $\alpha_{\rm B}$ determines the mixing of the kinetic contributions of the metric and scalar fields.
The decaying early modifications of gravity constrained here are therefore realized in a Horndeski scalar-tensor model with
\begin{equation}
 \alpha_{\rm X, model} = \left\{\begin{array}{ll} \alpha_{\rm X}, & z \leq z_{\rm on}, \\ 0, & z>z_{\rm on}, \end{array} \right.
\end{equation}
where the $\alpha_{\rm X}$ are given by Eq.~(\ref{eq:eft}) according to hybrid metric--Palatini gravity.
Note that $\alpha_{\rm X, model}(z>z_{\rm on})=0$ recovers a $\Lambda$CDM universe at high redshifts, avoiding the stringent high-curvature constraints around recombination.

Stability of the background solution of the Horndeski model with respect to the scalar mode requires~\cite{eftformalism}
\begin{equation}
Q_{\rm s} \equiv \frac{2 M_{*}^{2} D}{(2-\alpha_{\rm B})^2} > 0 \,,
\end{equation}
where
\begin{equation}
D \equiv \alpha_{\rm K} + \frac{3}{2}\alpha_{\rm B}^{2} = -\frac{3(f_{\mathcal{R}}')^2}{2f_{\mathcal{R}} (1+f_{\mathcal{R}})^2} \,.
\end{equation}
With the evolution of $f_{\mathcal{R}}$ given by hybrid metric-Palatini theory, we have
\begin{equation}
Q_{\rm s} = \left\{\begin{array}{ll} < 0\,, & {\rm for} \hspace{.9mm} f_{\mathcal{R}} > 0 \,, \\ > 0\,, & {\rm for} \hspace{.9mm} f_{\mathcal{R}} < 0\,. \end{array} \right.
\end{equation}
Hence, we require $-1<f_{\mathcal{R}}<0$ to prevent ghost instabilities.
To avoid a gradient instability or a superluminal sound speed $c_{\rm s}$ of the scalar field perturbation, we require that $0<c_{\rm s}^2\leq1$. To check this, we compute $c_{\rm s}^2$ in the hybrid metric-Palatini theory,
\begin{eqnarray}
D \cdot c_{\rm s}^2 & = & \frac{\kappa^2}{H^2 (1+f_{\mathcal{R}})}\left(\frac{4}{3}\rho_{\rm r} + \rho_{\rm m}\right)\left(f_{\mathcal{R}} + \frac{f_{\mathcal{R}}'}{2}\right) \nonumber\\
 & & + \frac{\alpha_{\rm M}}{2}\left(\frac{f_{\mathcal{R}}'+2(1+f_{\mathcal{R}})}{1+f_{\mathcal{R}}}\right) -\frac{f_{\mathcal{R}}''-(f_{\mathcal{R}}')^2}{1+f_{\mathcal{R}}} \,.
\end{eqnarray}
Furthermore, note that for the designer model we use in this work
\begin{equation}
 f_{\mathcal{R}}' = \left\{ \begin{array}{ll} >0\,, & {\rm for \ } f_{\mathcal{R}}<0 \,, \\ <0\,, & {\rm for \ } f_{\mathcal{R}}>0\,, \end{array} \right.
\end{equation}
and $|f_{\mathcal{R}}'| \gg |f_{\mathcal{R}}|$. Therefore, for $f_{\mathcal{R}}<0$, $f_{\mathcal{R}} + f_{\mathcal{R}}'/2 > 0$. Also, $f_{\mathcal{R}}^{\prime \prime}$ will be negative-definite (as can be verified by differentiating Eq.~(\ref{initialFprime})) for negative values of the scalar field. All of this, in conjunction with the fact that $\alpha_{\rm M}>0$ and $D>0$, ensures that $c_{\rm s}^2>0$ for $-1<f_{\mathcal{R}}<0$.
We have also confirmed numerically that $c_{\rm s}$ is subluminal for the range of values we consider for $f_{\mathcal{R}\rm{i}}$.
Note that whereas the condition for avoiding ghost instabilities applies to all hybrid metric-Palatini gravity models and should be respected when designing any other $f(\mathcal{R})$ models, the condition for avoiding gradient instabilities may be model dependent and should be studied in more detail for other choices of $f(\mathcal{R})$.
For completeness, we also verify the stability of tensor modes~\cite{eftformalism} with $Q_{\rm{T}} \propto \kappa^2 M^{2}_{\star}=1+f_{\mathcal{R}}>0$ whenever $f_{\mathcal{R}} > -1$. Also note that in $f(\mathcal{R})$ models, the propagation speed of gravitational waves equals the speed of light $c_{\rm{T}}=1$.

\section{Observational Constraints} \label{II}

Having fully specified a theoretically consistent decaying early modified gravity model in Sec.~\ref{sec:model}, we now determine the observational effects and constraints that can be set on early gravitational modifications with current cosmological data (Sec.~\ref{results}).
We also provide an outlook of constraints achievable with future surveys (Sec.~\ref{21cm}).
\subsection{Cosmological Observables} \label{survey}

To constrain our model parameters, we perform a MCMC search using
a range of geometric probes and CMB measurements by Planck~2015.

\subsubsection{Geometric Probes} \label{sec:geometricprobes}

The comparison between the luminosity magnitudes of high-redshift to low-redshift supernovae Type Ia (SNe~Ia) provides a relative distance measure affected by the Universe's expansion rate. Complementary absolute distance measures are obtained from measuring the local Hubble constant $H_0$ and the baryon acoustic oscillations (BAO) in the clustering of galaxies. These probes constrain the cosmological background evolution and since the $f(\mathcal{R})$ models considered here are designed to match the $\Lambda$CDM expansion history, they only serve to constrain the standard cosmological parameters and prevent degeneracies with the effect of the additional scalar degree of freedom on the fluctuations.

\subsubsection{Cosmic Microwave Background}\label{cmb}

In addition to the geometic probes described in Sec.~\ref{sec:geometricprobes}, the acoustic peaks in the CMB also contain information on the absolute distance to the last-scattering surface. These peaks are affected by early-time departures from GR at high curvature, i.e., in the case of $f(\mathcal{R})$ modifications, where $z_{\rm on}$ is sufficiently large.
Gravitational modifications can generally further manifest themselves in the CMB temperature and polarization via secondary anisotropies.
\begin{figure}[t!]
\begin{center}
\includegraphics[scale = 0.46]{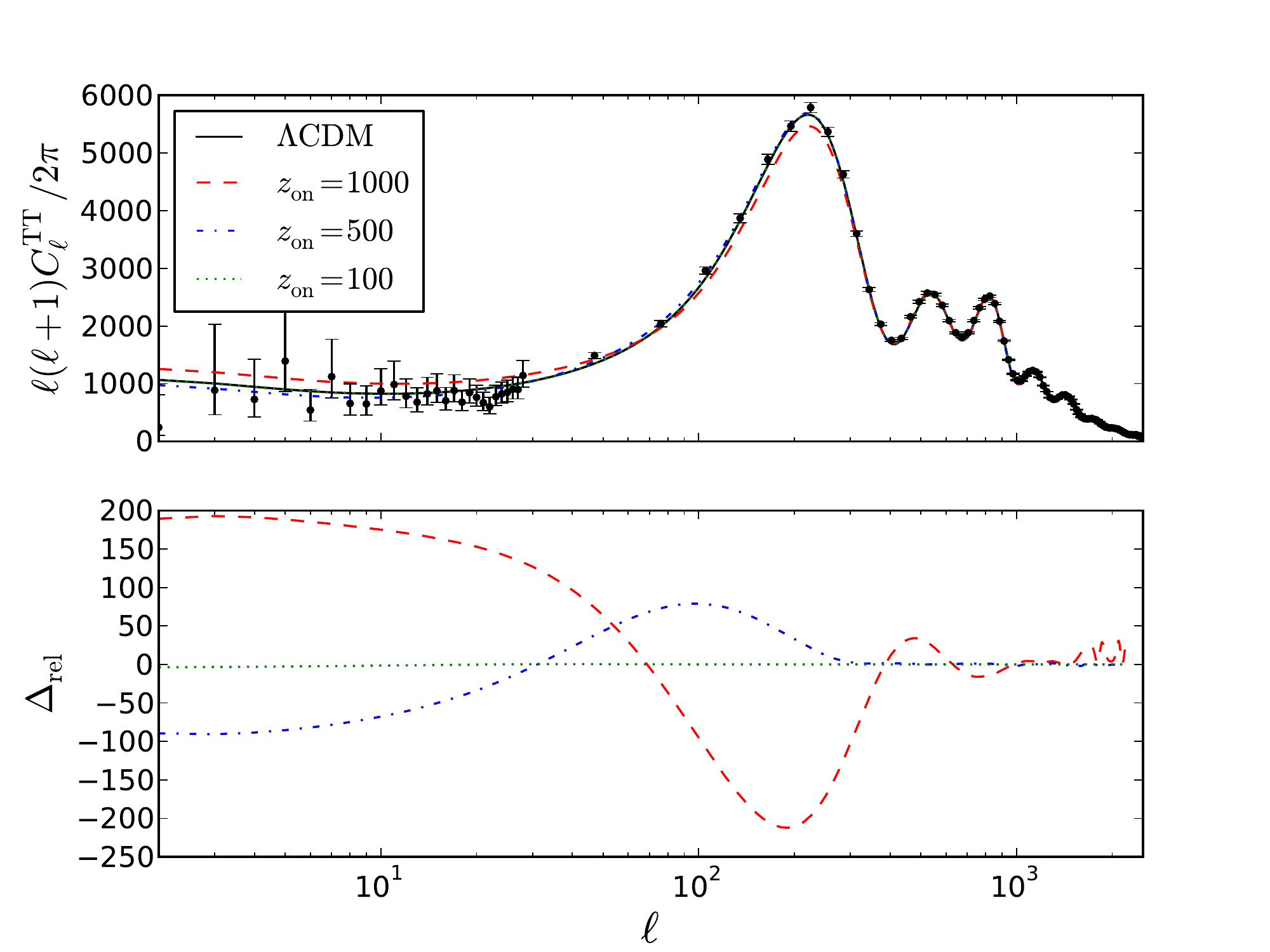}
\end{center}
\caption{\label{mgcambtt_evol} The lensed CMB temperature anisotropy power spectrum predicted by the designer hybrid metric-Palatini model for $|f_{\mathcal{R}}(z_{\rm{i}})| = 5 \times 10^{-2}$ and different values of $z_{\rm on}$ as well as the prediction for the $\Lambda$CDM model (top panel). The lower panel shows the difference to $\Lambda$CDM, $\Delta_{\rm{rel}} = \ell(\ell+1)\left( C_{\ell}^{TT, \rm{hybrid}} - C_{\ell}^{TT, \Lambda} \right)/(2\pi)$.}
\end{figure}
For details on the numerical computation of these effects in the designer hybrid metric-Palatini model, we refer the reader to Appendix~\ref{phenom}.

In Fig.~\ref{mgcambtt_evol}, we show the predictions for the CMB temperature anisotropy power spectrum (TT) for three different choices of $z_{\rm{on}}$. Hence, we introduce the oscillations between the Newtonian potentials in distinct epochs of the cosmological evolution which in turn produces different effects in the observed power spectrum. 
The first immediate observation is that, the later we introduce these oscillations, the less significant is their impact on the TT power spectrum. This is mainly due to the fact that, at later epochs, the amplitude of the oscillations have already been considerably damped out, reducing their effect on the TT power spectrum.

The second noticeable modification of the spectrum is in the Sachs-Wolfe plateau, on scales around $l<100$, where we observe a shift towards higher or smaller values compared to $\Lambda$CDM. The Sachs-Wolfe effect, resulting from a combination of gravitational redshift and intrinsic temperature fluctuations at angular last-scattering, can lead to a variation of the temperature power spectrum like \cite{sachswolfe} 
\begin{equation}{\label{sachswolfeeffect}}
 \frac{\Delta T}{T} \propto \delta \Phi,
\end{equation} 
where $\delta \Phi$ corresponds to the variation of the gravitational potential $\Phi$. The designer hybrid-metric Palatini model introduces modifications close to the surface of last-scattering. Therefore, depending on the redshift we choose to start the oscillations, the Newtonian potential $\Phi$ will be displaced toward larger or smaller values compared to $\Lambda$CDM, leading to the shift we observe in the power spectrum.
Then, at low $\ell$, we have the traditional increase in power due to the integrated Sachs-Wolfe (ISW) effect in the presence of late-time dark energy. Our model clearly mimics $\Lambda$CDM due to the fact that we fix the background evolution to match the standard cosmological scenario, even if the power can be deviated toward lower or smaller values due to the Sachs-Wolfe effect discussed before.

Lastly, we have what is probably the most discerning effect on the CMB TT power spectrum. When we introduce the oscillations at $z_{\rm{on}} = 1000$, we notice a significant decrease in the amplitude of the first peak. Traditionally, at early times, the non-negligible presence of radiation after the epoch of last-scattering can cause a decay of the gravitational potentials before these become constant, contributing to an early ISW effect that can influence the amplitude and position of the peaks. Therefore, if we allow modified gravity to be relevant close to the epoch of recombination, we not only modify this decay but also cause additional variation, influencing the acoustic phenomenology of the CMB. Of course, as we test lower valus of $z_{\rm{on}}$, this effect becomes increasingly negligible.

\subsection{\label{results}Cosmological Constraints}

Before presenting the current cosmological constraints on decaying early modified gravity, we briefly describe the cosmological datasets we use in our analysis.
We then give an outlook on constraints that can be obtained with 21~cm surveys and gravitational wave observations.

\subsubsection{Datasets} \label{sec:datasets}

For the SN~Ia luminosity-redshift relation, we use the dataset compiled in the Joint Lightcurve Analysis (JLA)~\cite{Betoule}. This includes records from the full three years of the Sloan Digital Sky Survey (SDSS) survey plus the ``C11 compilation'' assembled by Conley et al. (2011); comprising supernovae from the Supernovae Legacy Survey (SNLS), the Hubble Space Telescope (HST) and several nearby experiments. This whole sample consists of 740 SNe Ia. \\
\indent For $H_0$, we include information provided by the Wide Field Camera 3 (WFC3) on HST. The objective of this project was to determine the Hubble constant from optical and infrared observations of over 600 Cepheid variables in the host galaxies of 8 SNe~Ia, which provide the calibration for a magnitude-redshift relation based on 240 SNe~Ia~\cite{HST}. Hence, we use the gaussian prior of $H_0=73.8 \pm 2.4$ km $\rm{s}^{-1}$ $\rm{Mpc}^{-1}$.

We also use the BAO observations from the 6dF Galaxy Redshift Survey (6dFGRS) at low redshift $z_{\rm eff}=0.106$ \cite{BAO1}, as well as DR7 MGS  from SDSS at $z_{\rm eff}=0.15$, from the value-added galaxy catalogs hosted by NYU (NYU-VAGC) ~\cite{BAO2} and  the BAO signal from the Baryon Oscillation Spectroscopic Survey (BOSS) DR11 at $z_{\rm eff}=0.57$~\cite{BAO3}.

Lastly, we use the Planck 2015 data for the CMB. The Planck temperature and polarization and Planck lensing likelihood codes may be found in the Planck Legacy Archive~\cite{Planck}.

\subsubsection{Constraints}\label{constraints}

\begin{table}
 \begin{tabular}{|c|c|c|c|c|}
  \hline\hline
  $z_{\rm on}$ & ${\rm sgn}(f_{\mathcal{R}})$ & $|f_{\mathcal{R}\rm{i}}| \equiv |f_{\mathcal{R}}(z_{\rm{i}})|$ & $|f_{\mathcal{R}}(z_{\rm on})|$ & $|f_{\mathcal{R}}(z=0)|$ \\
  \hline
  1000 & $\pm$ & $<1.3\times10^{-2}$ & $<1.3\times10^{-2}$ & $<1.3 \times 10^{-8}$ \\
   500 & $\pm$ & $<4.7\times10^{-2}$ & $<1.2\times10^{-2}$ & $<4.7 \times 10^{-8}$ \\
   100 & $\pm$ & --- & --- & --- \\
  \hline
  1000 & $-$ & $<1.1\times10^{-2}$ & $<1.1\times10^{-2}$ & $<1.1\times10^{-8}$ \\
   500 & $-$ & $<4.8\times10^{-2}$ & $<1.2\times10^{-2}$ & $<4.8\times10^{-8}$ \\
   100 & $-$ & --- & --- & --- \\
  \hline
 \end{tabular}
 \caption{Current constraints (95\%~C.L.) on $f_{\mathcal{R}}(z_{\rm{i}}=1000)$ from the combination of surveys discussed in Sec.~\ref{sec:datasets}. Note that models with a positive sign of $f_{\mathcal{R}}$ suffer from a ghost instability (see Sec.~\ref{coneft}) and models with $z_{\rm on}=100$ cannot be constrained within the prior $|f_{\mathcal{R}\rm{i}}|<0.1$ required for the viability of the approximations performed in Sec.~\ref{earlyapp}. However, a constraint of $|f_{\mathcal{R}\rm{i}}|\lesssim10^{-3}$ on all models will be achievable with 21~cm intensity mapping (see Sec.~\ref{21cm}). We also present constraints on the value of $f_{\mathcal{R}}$ at the redshift of decoupling, $z_{\rm{on}}$, and at the present time, $z = 0$.
 \label{tab:constraints}}
\end{table}

Using the datasets described in Sec.~\ref{sec:datasets}, we conduct an MCMC parameter estimation analysis with {\sc cosmomc}~\cite{cosmomc} (see Appendix~\ref{phenom} for details).
We summarize our constraints on the early-time decaying modified gravity model of Sec.~\ref{sec:model} in Table~\ref{tab:constraints}.
It is easily noticeable that the constraining power of the data over the model changes significantly the later we introduce the oscillations between the Newtonian potentials ($z \leq z_{\rm on}$).

For $z_{\rm{on}} = 1000$, allowing both signs for $f_{\mathcal{R}{\rm i}} \equiv f_{\mathcal{R}}(z_{\rm{i}} = 1000)$, we infer a 1D-marginalized constraint of
$|f_{\mathcal{R}\rm{i}}|<1.3 \times 10^{-2}$ (95\% C.L.), where we adopt a flat symmetric prior $f_{\mathcal{R}{\rm{i}}} \in [-0.1,0.1]$.
We stress, however, that positive values of $f_{\mathcal{R}\rm{i}}$ are affected by the ghost instability discussed in Sec.~\ref{coneft}. Considering the stable branch only with a negative flat prior, we find $|f_{\mathcal{R}\rm{i}}| < 1.1 \times 10^{-2}$.
These values are comparable to the constraints obtained in Ref.~\cite{hybridback} on $f(\mathcal{R})$ models that deviate from the $\Lambda$CDM expansion history, using background data alone. Although we note that these constraints have been inferred for initial modifications at much higher redshift.
$\Lambda$CDM is clearly the favored model and we find no evidence for early-time modifications in the observations.
The constraints we found are mostly driven by two prominent effects on the CMB that we have observed in Sec.~\ref{cmb}: a modification of the Sachs-Wolfe plateau and of the amplitude of the first peak. However, there is also a non-negligible contribution of CMB lensing, which is sensitive to percent-level modifications at high $\ell$~\cite{cmblens} and can constrain the effects of $z_{\rm{on}}=1000$ shown
in Fig.~\ref{mgcambtt_evol}.
We also note that the present absolute value of the scalar field, $|f_{\mathcal{R}0}| \equiv |f_{\mathcal{R}}(z=0)|$, is very small and of order $10^{-8}$. This implies that modifications are strongly suppressed at the smallest scales, where these are proportional to the background value of the scalar field~\cite{main1} (see Sec.~\ref{perturbations}).

Finally, decreasing $z_{\rm{on}}$ leads to a considerable weakening of the constraints on the early-time deviation from GR.
With $z_{\rm{on}} = 500$, constraints on the scalar field value at equal redshift weaken by a factor of approximately $4$.
For $z_{\rm{on}} = 100$, we can no longer constrain the scalar field value within the prior $|f_{\mathcal{R}{\rm i}}|<0.1$.
This is due to the oscillations on the slip between the gravitational potentials being significantly damped out by $z=100$, hence only introducing very small deviations from GR.

\subsection{Outlook: 21~cm and Gravitational Waves}\label{21cm}

Finally, we provide rough estimates of the constraints on early decaying modified gravity that will be achievable with 21~cm intensity mapping~\cite{madau:97,brax:12,hall:12} and standard sirens~\cite{schutz:86,holz:05,lombriser:15b} using gravitational waves emitted by events at cosmological distances.
To estimate constraints obtainable with 21~cm surveys, we compare deviations in the matter power spectrum between the model and $\Lambda$CDM to bounds on modified gravity reported in Ref.~\cite{brax:12} at $z=11$ and Ref.~\cite{hall:12} at $z=2.5$.
We find that $|f_{\mathcal{R}\rm{i}}|\lesssim10^{-3}$ and $|f_{\mathcal{R}\rm{i}}|\lesssim5\times10^{-2}$ for
$z_{\rm on}=1000$,
which is competitive with the CMB constraints in Table~\ref{tab:constraints}.
Standard sirens will constrain the luminosity distance at $z\sim(1-2)$ at the $\sim1\%$ level, and at the $\sim10\%$ level for $z\sim7$~\cite{cutler:09,tamanini:16}.
In modified gravity models, this constraint can be used to set a bound on the evolution of the Planck mass~\cite{lombriser:15b}, which for our model corresponds to a constraint of $|f_{\mathcal{R}\rm{i}}|\lesssim10^3$, which will not be competitive with the constraints in Table~\ref{tab:constraints}.

\section{\label{III}Conclusions}

In this work we have explored the current cosmological constraints that can be inferred on modifications of gravity which may become significant at early times after recombination and decay towards the present.
We have chosen the designer hybrid metric-Palatini model as a specific example of an early-time modification of gravity. Fixing the background evolution to exactly match $\Lambda$CDM, we are able to separate background constraints from constraints inferred from the modified dynamics of linear perturbations due to the impact that these have on the CMB. We also describe how this model can be realized in the more general context of the effective field theory formalism of Horndeski gravity, and study its stability. We conclude that the model is stable as long as the additional scalar degree of freedom introduced by the hybrid metric-Palatini theory remains negative with an amplitude smaller than unity, which 
implies an effective enhancement of the gravitational coupling.

In order to perform efficient numerical computations, we have developed an approximation for the evolution of the slip between the Newtonian potentials that is valid beyond the standard quasistatic subhorizon approximation.
This extension becomes important at high redshifts, where we show that a quasistatic approach alone breaks down due to the known oscillations of the linear perturbations of the model~\cite{hybridpert}.

Using a combination of observational data on the background evolution and of the CMB anisotropies, we infer constraints on the allowed early-time deviations from GR.
The results we obtain are dependent on the redshift at which we introduce the oscillations in the slip between the gravitational potentials. If these are set at $z_{\rm{on}} = 1000$, we are able to constrain the absolute deviation from GR at $z_{\rm on}$ to $\lesssim 10^{-2}$ at the $95\%$ confidence level. This result is comparable to the constraints obtained from background data alone in Ref.~\cite{hybridback} for $f(\mathcal{R})$ models that depart from the $\Lambda$CDM expansion history.

The constraints we obtain at this redshift can be attributed to noticeable effects on the CMB power spectrum. We are able to observe a substantial shift in the Sachs-Wolfe plateau due to a modification of the Newtonian potential $\Phi$ at a time close to recombination. There is also a significant suppression of the first peak due to complementary variation of the gravitational potentials close to the epoch of recombination that, together with the non-negligible presence of radiation, contributes to an early integrated Sachs-Wolfe effect that can alter the amplitude and position of the peaks. Smaller contributions to the constraints can be attributed to CMB lensing which is sensitive to the percent-level modifications we observe at high $\ell$.
Finally, we find that future 21~cm survey data will significantly improve upon the CMB constraints, whereas using gravitational wave events as standard sirens will not provide competitive bounds.

\vspace{5 mm}
\acknowledgments

We thank Andrew Liddle, Alex Hall and Tomi Koivisto for useful discussions and comments on this manuscript. N.A.L.\ acknowledges financial support from Funda\c{c}\~{a}o para a Ci\^{e}ncia e a Tecnologia (FCT) through grant SFRH/BD/85164/2012.  V.S.-B.\ acknowledges funding provided by CONACyT and the University of Edinburgh. L.L.\ was supported by the STFC Consolidated Grant for Astronomy and Astrophysics at the University of Edinburgh and a SNSF Advanced Postdoc.Mobility Fellowship (No.~161058).
Numerical computations were conducted on the COSMOS Shared Memory system at DAMTP, University of Cambridge operated on behalf of the STFC DiRAC HPC Facility. This equipment is funded by BIS National E-infrastructure capital grant ST/J005673/1 and STFC grants ST/H008586/1, ST/K00333X/1.

\bibliography{hybrid_cmb}

\begin{thebibliography}{61}%
\makeatletter
\providecommand \@ifxundefined [1]{%
 \@ifx{#1\undefined}
}%
\providecommand \@ifnum [1]{%
 \ifnum #1\expandafter \@firstoftwo
 \else \expandafter \@secondoftwo
 \fi
}%
\providecommand \@ifx [1]{%
 \ifx #1\expandafter \@firstoftwo
 \else \expandafter \@secondoftwo
 \fi
}%
\providecommand \natexlab [1]{#1}%
\providecommand \enquote  [1]{``#1''}%
\providecommand \bibnamefont  [1]{#1}%
\providecommand \bibfnamefont [1]{#1}%
\providecommand \citenamefont [1]{#1}%
\providecommand \href@noop [0]{\@secondoftwo}%
\providecommand \href [0]{\begingroup \@sanitize@url \@href}%
\providecommand \@href[1]{\@@startlink{#1}\@@href}%
\providecommand \@@href[1]{\endgroup#1\@@endlink}%
\providecommand \@sanitize@url [0]{\catcode `\\12\catcode `\$12\catcode
  `\&12\catcode `\#12\catcode `\^12\catcode `\_12\catcode `\%12\relax}%
\providecommand \@@startlink[1]{}%
\providecommand \@@endlink[0]{}%
\providecommand \url  [0]{\begingroup\@sanitize@url \@url }%
\providecommand \@url [1]{\endgroup\@href {#1}{\urlprefix }}%
\providecommand \urlprefix  [0]{URL }%
\providecommand \Eprint [0]{\href }%
\providecommand \doibase [0]{http://dx.doi.org/}%
\providecommand \selectlanguage [0]{\@gobble}%
\providecommand \bibinfo  [0]{\@secondoftwo}%
\providecommand \bibfield  [0]{\@secondoftwo}%
\providecommand \translation [1]{[#1]}%
\providecommand \BibitemOpen [0]{}%
\providecommand \bibitemStop [0]{}%
\providecommand \bibitemNoStop [0]{.\EOS\space}%
\providecommand \EOS [0]{\spacefactor3000\relax}%
\providecommand \BibitemShut  [1]{\csname bibitem#1\endcsname}%
\let\auto@bib@innerbib\@empty
%</preamble>
\bibitem [{\citenamefont {Clifton}\ \emph {et~al.}(2012)\citenamefont
  {Clifton}, \citenamefont {Ferreira}, \citenamefont {Padilla},\ and\
  \citenamefont {Skordis}}]{reviewall}%
  \BibitemOpen
  \bibfield  {author} {\bibinfo {author} {\bibfnamefont {T.}~\bibnamefont
  {Clifton}}, \bibinfo {author} {\bibfnamefont {P.~G.}\ \bibnamefont
  {Ferreira}}, \bibinfo {author} {\bibfnamefont {A.}~\bibnamefont {Padilla}}, \
  and\ \bibinfo {author} {\bibfnamefont {C.}~\bibnamefont {Skordis}},\
  }\href@noop {} {\bibfield  {journal} {\bibinfo  {journal} {Phys. Rept.}\
  }\textbf {\bibinfo {volume} {513}},\ \bibinfo {pages} {1} (\bibinfo {year}
  {2012})},\ \Eprint {http://arxiv.org/abs/arXiv:1106.2476v3}
  {arXiv:1106.2476v3} \BibitemShut {NoStop}%
\bibitem [{\citenamefont {Koyama}(2016)}]{koyama:15}%
  \BibitemOpen
  \bibfield  {author} {\bibinfo {author} {\bibfnamefont {K.}~\bibnamefont
  {Koyama}},\ }\href@noop {} {\bibfield  {journal} {\bibinfo  {journal} {Rept.
  Prog. Phys.}\ }\textbf {\bibinfo {volume} {79}},\ \bibinfo {pages} {046902}
  (\bibinfo {year} {2016})},\ \Eprint {http://arxiv.org/abs/1504.04623}
  {arXiv:1504.04623 [astro-ph.CO]} \BibitemShut {NoStop}%
%%CITATION = ARXIV:1504.04623;%%
\bibitem [{\citenamefont {Joyce}\ \emph {et~al.}(2016)\citenamefont {Joyce},
  \citenamefont {Lombriser},\ and\ \citenamefont {Schmidt}}]{joyce:16}%
  \BibitemOpen
  \bibfield  {author} {\bibinfo {author} {\bibfnamefont {A.}~\bibnamefont
  {Joyce}}, \bibinfo {author} {\bibfnamefont {L.}~\bibnamefont {Lombriser}}, \
  and\ \bibinfo {author} {\bibfnamefont {F.}~\bibnamefont {Schmidt}},\
  }\href@noop {} {\bibfield  {journal} {\bibinfo  {journal} {Annu. Rev. Nucl.
  Part. Sci}\ }\textbf {\bibinfo {volume} {66}},\ \bibinfo {pages} {95}
  (\bibinfo {year} {2016})},\ \Eprint {http://arxiv.org/abs/1601.06133}
  {arXiv:1601.06133 [astro-ph.CO]} \BibitemShut {NoStop}%
%%CITATION = ARXIV:1601.06133;%%
\bibitem [{\citenamefont {Abbott}\ \emph {et~al.}(2015)\citenamefont {Abbott}
  \emph {et~al.}}]{des}%
  \BibitemOpen
  \bibfield  {author} {\bibinfo {author} {\bibfnamefont {T.}~\bibnamefont
  {Abbott}} \emph {et~al.} (\bibinfo {collaboration} {The Dark Energy Survey
  Collaboration}),\ }\href@noop {} {\  (\bibinfo {year} {2015})},\ \Eprint
  {http://arxiv.org/abs/arXiv:astro-ph/0510346} {arXiv:astro-ph/0510346}
  \BibitemShut {NoStop}%
\bibitem [{\citenamefont {Dawson}\ \emph {et~al.}(2016)\citenamefont {Dawson}
  \emph {et~al.}}]{eboss}%
  \BibitemOpen
  \bibfield  {author} {\bibinfo {author} {\bibfnamefont {K.~S.}\ \bibnamefont
  {Dawson}} \emph {et~al.},\ }\href@noop {} {\bibfield  {journal} {\bibinfo
  {journal} {The Astronomical Journal}\ }\textbf {\bibinfo {volume} {151}},\
  \bibinfo {pages} {44} (\bibinfo {year} {2016})},\ \Eprint
  {http://arxiv.org/abs/arXiv:1508.04473} {arXiv:1508.04473} \BibitemShut
  {NoStop}%
\bibitem [{\citenamefont {Laureijs}\ \emph {et~al.}(2011)\citenamefont
  {Laureijs} \emph {et~al.}}]{euclid}%
  \BibitemOpen
  \bibfield  {author} {\bibinfo {author} {\bibfnamefont {R.}~\bibnamefont
  {Laureijs}} \emph {et~al.},\ }\href@noop {} {\bibfield  {journal} {\bibinfo
  {journal} {ESA/SRE(2011)12}\ } (\bibinfo {year} {2011})},\ \Eprint
  {http://arxiv.org/abs/arXiv:1110.3193} {arXiv:1110.3193} \BibitemShut
  {NoStop}%
\bibitem [{\citenamefont {Riess}\ \emph {et~al.}(1998)\citenamefont {Riess}
  \emph {et~al.}}]{accel1}%
  \BibitemOpen
  \bibfield  {author} {\bibinfo {author} {\bibfnamefont {A.~G.}\ \bibnamefont
  {Riess}} \emph {et~al.},\ }\href@noop {} {\bibfield  {journal} {\bibinfo
  {journal} {Astron. J.}\ }\textbf {\bibinfo {volume} {116}},\ \bibinfo {pages}
  {1009} (\bibinfo {year} {1998})},\ \Eprint
  {http://arxiv.org/abs/astro-ph/9805201} {astro-ph/9805201} \BibitemShut
  {NoStop}%
\bibitem [{\citenamefont {Perlmutter}\ \emph {et~al.}(1999)\citenamefont
  {Perlmutter} \emph {et~al.}}]{accel2}%
  \BibitemOpen
  \bibfield  {author} {\bibinfo {author} {\bibfnamefont {S.}~\bibnamefont
  {Perlmutter}} \emph {et~al.},\ }\href@noop {} {\bibfield  {journal} {\bibinfo
   {journal} {Astrophy. J.}\ }\textbf {\bibinfo {volume} {517}},\ \bibinfo
  {pages} {565} (\bibinfo {year} {1999})},\ \Eprint
  {http://arxiv.org/abs/astro-ph/9812133} {astro-ph/9812133} \BibitemShut
  {NoStop}%
\bibitem [{\citenamefont {Riess}\ \emph {et~al.}(2007)\citenamefont {Riess}
  \emph {et~al.}}]{accel3}%
  \BibitemOpen
  \bibfield  {author} {\bibinfo {author} {\bibfnamefont {A.~G.}\ \bibnamefont
  {Riess}} \emph {et~al.},\ }\href@noop {} {\bibfield  {journal} {\bibinfo
  {journal} {Astroph. J.}\ }\textbf {\bibinfo {volume} {659}},\ \bibinfo
  {pages} {98} (\bibinfo {year} {2007})},\ \Eprint
  {http://arxiv.org/abs/astro-ph/0612666} {astro-ph/0612666} \BibitemShut
  {NoStop}%
\bibitem [{\citenamefont {Amanullah}\ \emph {et~al.}(2010)\citenamefont
  {Amanullah} \emph {et~al.}}]{accel4}%
  \BibitemOpen
  \bibfield  {author} {\bibinfo {author} {\bibfnamefont {R.}~\bibnamefont
  {Amanullah}} \emph {et~al.},\ }\href@noop {} {\bibfield  {journal} {\bibinfo
  {journal} {Astroph. J.}\ }\textbf {\bibinfo {volume} {716}},\ \bibinfo
  {pages} {712} (\bibinfo {year} {2010})},\ \Eprint
  {http://arxiv.org/abs/arXiv:1004.1711} {arXiv:1004.1711} \BibitemShut
  {NoStop}%
\bibitem [{\citenamefont {Padmanabhan}(2003)}]{lambdareview}%
  \BibitemOpen
  \bibfield  {author} {\bibinfo {author} {\bibfnamefont {T.}~\bibnamefont
  {Padmanabhan}},\ }\href@noop {} {\bibfield  {journal} {\bibinfo  {journal}
  {Phys. Rept.}\ }\textbf {\bibinfo {volume} {380}},\ \bibinfo {pages} {235}
  (\bibinfo {year} {2003})},\ \Eprint {http://arxiv.org/abs/hep-th/0212290}
  {hep-th/0212290} \BibitemShut {NoStop}%
\bibitem [{\citenamefont {Lombriser}\ and\ \citenamefont
  {Lima}(2016)}]{lombriser:16}%
  \BibitemOpen
  \bibfield  {author} {\bibinfo {author} {\bibfnamefont {L.}~\bibnamefont
  {Lombriser}}\ and\ \bibinfo {author} {\bibfnamefont {N.~A.}\ \bibnamefont
  {Lima}},\ }\href@noop {} {\  (\bibinfo {year} {2016})},\ \Eprint
  {http://arxiv.org/abs/1602.07670} {arXiv:1602.07670 [astro-ph.CO]}
  \BibitemShut {NoStop}%
%%CITATION = ARXIV:1602.07670;%%
\bibitem [{\citenamefont {Brans}\ and\ \citenamefont {Dicke}(1961)}]{bd}%
  \BibitemOpen
  \bibfield  {author} {\bibinfo {author} {\bibfnamefont {C.~H.}\ \bibnamefont
  {Brans}}\ and\ \bibinfo {author} {\bibfnamefont {R.~H.}\ \bibnamefont
  {Dicke}},\ }\href@noop {} {\bibfield  {journal} {\bibinfo  {journal} {Phys.
  Rev.}\ }\textbf {\bibinfo {volume} {124}},\ \bibinfo {pages} {925} (\bibinfo
  {year} {1961})}\BibitemShut {NoStop}%
\bibitem [{\citenamefont {Nicolis}\ \emph {et~al.}(2009)\citenamefont
  {Nicolis}, \citenamefont {Rattazzi},\ and\ \citenamefont
  {Trincherini}}]{Gall}%
  \BibitemOpen
  \bibfield  {author} {\bibinfo {author} {\bibfnamefont {A.}~\bibnamefont
  {Nicolis}}, \bibinfo {author} {\bibfnamefont {R.}~\bibnamefont {Rattazzi}}, \
  and\ \bibinfo {author} {\bibfnamefont {E.}~\bibnamefont {Trincherini}},\
  }\href@noop {} {\bibfield  {journal} {\bibinfo  {journal} {Phys. Rev. D}\
  }\textbf {\bibinfo {volume} {79}},\ \bibinfo {pages} {064036} (\bibinfo
  {year} {2009})},\ \Eprint {http://arxiv.org/abs/arXiv:0811.2197v2}
  {arXiv:0811.2197v2} \BibitemShut {NoStop}%
\bibitem [{\citenamefont {Sotiriou}\ and\ \citenamefont
  {Faraoni}(2010)}]{frreview}%
  \BibitemOpen
  \bibfield  {author} {\bibinfo {author} {\bibfnamefont {T.~P.}\ \bibnamefont
  {Sotiriou}}\ and\ \bibinfo {author} {\bibfnamefont {V.}~\bibnamefont
  {Faraoni}},\ }\href@noop {} {\bibfield  {journal} {\bibinfo  {journal} {Rev.
  Mod. Phys.}\ }\textbf {\bibinfo {volume} {82}},\ \bibinfo {pages} {451}
  (\bibinfo {year} {2010})},\ \Eprint {http://arxiv.org/abs/arXiv:0805.1726v4}
  {arXiv:0805.1726v4} \BibitemShut {NoStop}%
\bibitem [{\citenamefont {Horndeski}(1974)}]{horndeski:74}%
  \BibitemOpen
  \bibfield  {author} {\bibinfo {author} {\bibfnamefont {G.~W.}\ \bibnamefont
  {Horndeski}},\ }\href {\doibase 10.1007/BF01807638} {\bibfield  {journal}
  {\bibinfo  {journal} {Int.J.Theor.Phys.}\ }\textbf {\bibinfo {volume} {10}},\
  \bibinfo {pages} {363} (\bibinfo {year} {1974})}\BibitemShut {NoStop}%
%%CITATION = IJTPB,10,363;%%
\bibitem [{\citenamefont {Lombriser}\ and\ \citenamefont
  {Taylor}(2016)}]{lombriser:15b}%
  \BibitemOpen
  \bibfield  {author} {\bibinfo {author} {\bibfnamefont {L.}~\bibnamefont
  {Lombriser}}\ and\ \bibinfo {author} {\bibfnamefont {A.}~\bibnamefont
  {Taylor}},\ }\href {\doibase 10.1088/1475-7516/2016/03/031} {\bibfield
  {journal} {\bibinfo  {journal} {JCAP}\ }\textbf {\bibinfo {volume} {1603}},\
  \bibinfo {pages} {031} (\bibinfo {year} {2016})},\ \Eprint
  {http://arxiv.org/abs/1509.08458} {arXiv:1509.08458 [astro-ph.CO]}
  \BibitemShut {NoStop}%
%%CITATION = ARXIV:1509.08458;%%
\bibitem [{\citenamefont {Lombriser}(2011)}]{lombriser:11}%
  \BibitemOpen
  \bibfield  {author} {\bibinfo {author} {\bibfnamefont {L.}~\bibnamefont
  {Lombriser}},\ }\href {\doibase 10.1103/PhysRevD.83.063519} {\bibfield
  {journal} {\bibinfo  {journal} {Phys. Rev.}\ }\textbf {\bibinfo {volume}
  {D83}},\ \bibinfo {pages} {063519} (\bibinfo {year} {2011})},\ \Eprint
  {http://arxiv.org/abs/1101.0594} {arXiv:1101.0594 [astro-ph.CO]} \BibitemShut
  {NoStop}%
%%CITATION = ARXIV:1101.0594;%%
\bibitem [{\citenamefont {Brax}\ \emph {et~al.}(2014)\citenamefont {Brax},
  \citenamefont {van~de Bruck}, \citenamefont {Clesse}, \citenamefont {Davis},\
  and\ \citenamefont {Sculthorpe}}]{brax:13}%
  \BibitemOpen
  \bibfield  {author} {\bibinfo {author} {\bibfnamefont {P.}~\bibnamefont
  {Brax}}, \bibinfo {author} {\bibfnamefont {C.}~\bibnamefont {van~de Bruck}},
  \bibinfo {author} {\bibfnamefont {S.}~\bibnamefont {Clesse}}, \bibinfo
  {author} {\bibfnamefont {A.-C.}\ \bibnamefont {Davis}}, \ and\ \bibinfo
  {author} {\bibfnamefont {G.}~\bibnamefont {Sculthorpe}},\ }\href {\doibase
  10.1103/PhysRevD.89.123507} {\bibfield  {journal} {\bibinfo  {journal} {Phys.
  Rev.}\ }\textbf {\bibinfo {volume} {D89}},\ \bibinfo {pages} {123507}
  (\bibinfo {year} {2014})},\ \Eprint {http://arxiv.org/abs/1312.3361}
  {arXiv:1312.3361 [astro-ph.CO]} \BibitemShut {NoStop}%
%%CITATION = ARXIV:1312.3361;%%
\bibitem [{\citenamefont {Kohri}\ \emph {et~al.}(2016)\citenamefont {Kohri},
  \citenamefont {Oyama}, \citenamefont {Sekiguchi},\ and\ \citenamefont
  {Takahashi}}]{21cmdark}%
  \BibitemOpen
  \bibfield  {author} {\bibinfo {author} {\bibfnamefont {K.}~\bibnamefont
  {Kohri}}, \bibinfo {author} {\bibfnamefont {Y.}~\bibnamefont {Oyama}},
  \bibinfo {author} {\bibfnamefont {T.}~\bibnamefont {Sekiguchi}}, \ and\
  \bibinfo {author} {\bibfnamefont {T.}~\bibnamefont {Takahashi}},\ }\href@noop
  {} {\  (\bibinfo {year} {2016})},\ \Eprint
  {http://arxiv.org/abs/arXiv:1608.01601} {arXiv:1608.01601} \BibitemShut
  {NoStop}%
\bibitem [{\citenamefont {Santos}\ \emph {et~al.}(2015)\citenamefont {Santos}
  \emph {et~al.}}]{ska}%
  \BibitemOpen
  \bibfield  {author} {\bibinfo {author} {\bibfnamefont {M.~G.}\ \bibnamefont
  {Santos}} \emph {et~al.},\ }\href@noop {} {\  (\bibinfo {year} {2015})},\
  \Eprint {http://arxiv.org/abs/arXiv:1501.03989} {arXiv:1501.03989}
  \BibitemShut {NoStop}%
\bibitem [{\citenamefont {Gleyzes}\ \emph {et~al.}(2014)\citenamefont
  {Gleyzes}, \citenamefont {Langlois},\ and\ \citenamefont
  {Vernizzi}}]{gleyzes:14}%
  \BibitemOpen
  \bibfield  {author} {\bibinfo {author} {\bibfnamefont {J.}~\bibnamefont
  {Gleyzes}}, \bibinfo {author} {\bibfnamefont {D.}~\bibnamefont {Langlois}}, \
  and\ \bibinfo {author} {\bibfnamefont {F.}~\bibnamefont {Vernizzi}},\ }\href
  {\doibase 10.1142/S021827181443010X} {\bibfield  {journal} {\bibinfo
  {journal} {Int. J. Mod. Phys.}\ }\textbf {\bibinfo {volume} {D23}},\ \bibinfo
  {pages} {3010} (\bibinfo {year} {2014})},\ \Eprint
  {http://arxiv.org/abs/1411.3712} {arXiv:1411.3712 [hep-th]} \BibitemShut
  {NoStop}%
%%CITATION = ARXIV:1411.3712;%%
\bibitem [{\citenamefont {Pogosian}\ and\ \citenamefont
  {Silvestri}(2008)}]{metricfrpert}%
  \BibitemOpen
  \bibfield  {author} {\bibinfo {author} {\bibfnamefont {L.}~\bibnamefont
  {Pogosian}}\ and\ \bibinfo {author} {\bibfnamefont {A.}~\bibnamefont
  {Silvestri}},\ }\href@noop {} {\bibfield  {journal} {\bibinfo  {journal}
  {Phys. Rev. D}\ }\textbf {\bibinfo {volume} {77}},\ \bibinfo {pages} {023503}
  (\bibinfo {year} {2008})},\ \Eprint {http://arxiv.org/abs/arXiv:0709.0296}
  {arXiv:0709.0296} \BibitemShut {NoStop}%
\bibitem [{\citenamefont {Jain}\ and\ \citenamefont
  {VanderPlas}(2011)}]{dwarf_galaxies}%
  \BibitemOpen
  \bibfield  {author} {\bibinfo {author} {\bibfnamefont {B.}~\bibnamefont
  {Jain}}\ and\ \bibinfo {author} {\bibfnamefont {J.}~\bibnamefont
  {VanderPlas}},\ }\href@noop {} {\bibfield  {journal} {\bibinfo  {journal}
  {JCAP}\ }\textbf {\bibinfo {volume} {2011}},\ \bibinfo {pages} {32} (\bibinfo
  {year} {2011})},\ \Eprint {http://arxiv.org/abs/arXiv:1106.0065}
  {arXiv:1106.0065} \BibitemShut {NoStop}%
\bibitem [{\citenamefont {Lombriser}\ \emph {et~al.}(2015)\citenamefont
  {Lombriser}, \citenamefont {Simpson},\ and\ \citenamefont
  {Mead}}]{lombriser:15}%
  \BibitemOpen
  \bibfield  {author} {\bibinfo {author} {\bibfnamefont {L.}~\bibnamefont
  {Lombriser}}, \bibinfo {author} {\bibfnamefont {F.}~\bibnamefont {Simpson}},
  \ and\ \bibinfo {author} {\bibfnamefont {A.}~\bibnamefont {Mead}},\ }\href
  {\doibase 10.1103/PhysRevLett.114.251101} {\bibfield  {journal} {\bibinfo
  {journal} {Phys. Rev. Lett.}\ }\textbf {\bibinfo {volume} {114}},\ \bibinfo
  {pages} {251101} (\bibinfo {year} {2015})},\ \Eprint
  {http://arxiv.org/abs/1501.04961} {arXiv:1501.04961 [astro-ph.CO]}
  \BibitemShut {NoStop}%
%%CITATION = ARXIV:1501.04961;%%
\bibitem [{\citenamefont {Harko}\ \emph {et~al.}(2012)\citenamefont {Harko},
  \citenamefont {Koivisto}, \citenamefont {Lobo},\ and\ \citenamefont
  {Olmo}}]{main1}%
  \BibitemOpen
  \bibfield  {author} {\bibinfo {author} {\bibfnamefont {T.}~\bibnamefont
  {Harko}}, \bibinfo {author} {\bibfnamefont {T.~S.}\ \bibnamefont {Koivisto}},
  \bibinfo {author} {\bibfnamefont {F.~S.~N.}\ \bibnamefont {Lobo}}, \ and\
  \bibinfo {author} {\bibfnamefont {G.~J.}\ \bibnamefont {Olmo}},\ }\href@noop
  {} {\bibfield  {journal} {\bibinfo  {journal} {Phys. Rev. D}\ }\textbf
  {\bibinfo {volume} {85}},\ \bibinfo {pages} {084016} (\bibinfo {year}
  {2012})},\ \Eprint {http://arxiv.org/abs/arXiv:1110.1049v2}
  {arXiv:1110.1049v2} \BibitemShut {NoStop}%
\bibitem [{\citenamefont {Capozziello}\ \emph
  {et~al.}(2013{\natexlab{a}})\citenamefont {Capozziello}, \citenamefont
  {Harko}, \citenamefont {Lobo},\ and\ \citenamefont {Olmo}}]{main2}%
  \BibitemOpen
  \bibfield  {author} {\bibinfo {author} {\bibfnamefont {S.}~\bibnamefont
  {Capozziello}}, \bibinfo {author} {\bibfnamefont {T.}~\bibnamefont {Harko}},
  \bibinfo {author} {\bibfnamefont {F.~S.~N.}\ \bibnamefont {Lobo}}, \ and\
  \bibinfo {author} {\bibfnamefont {G.~J.}\ \bibnamefont {Olmo}},\ }\href@noop
  {} {\bibfield  {journal} {\bibinfo  {journal} {Int. J. Mod. Phys. D}\
  }\textbf {\bibinfo {volume} {22}},\ \bibinfo {pages} {1342006} (\bibinfo
  {year} {2013}{\natexlab{a}})},\ \Eprint
  {http://arxiv.org/abs/arXiv:1305.3756v2} {arXiv:1305.3756v2} \BibitemShut
  {NoStop}%
\bibitem [{\citenamefont {Capozziello}\ \emph {et~al.}(2015)\citenamefont
  {Capozziello}, \citenamefont {Harko}, \citenamefont {Koivisto}, \citenamefont
  {Lobo},\ and\ \citenamefont {Olmo}}]{Capozziello:2015lza}%
  \BibitemOpen
  \bibfield  {author} {\bibinfo {author} {\bibfnamefont {S.}~\bibnamefont
  {Capozziello}}, \bibinfo {author} {\bibfnamefont {T.}~\bibnamefont {Harko}},
  \bibinfo {author} {\bibfnamefont {T.~S.}\ \bibnamefont {Koivisto}}, \bibinfo
  {author} {\bibfnamefont {F.~S.~N.}\ \bibnamefont {Lobo}}, \ and\ \bibinfo
  {author} {\bibfnamefont {G.~J.}\ \bibnamefont {Olmo}},\ }\href {\doibase
  10.3390/universe1020199} {\bibfield  {journal} {\bibinfo  {journal}
  {Universe}\ }\textbf {\bibinfo {volume} {1}},\ \bibinfo {pages} {199}
  (\bibinfo {year} {2015})},\ \Eprint {http://arxiv.org/abs/1508.04641}
  {arXiv:1508.04641 [gr-qc]} \BibitemShut {NoStop}%
%%CITATION = ARXIV:1508.04641;%%
\bibitem [{\citenamefont {Olmo}(2011)}]{palainst1}%
  \BibitemOpen
  \bibfield  {author} {\bibinfo {author} {\bibfnamefont {G.~J.}\ \bibnamefont
  {Olmo}},\ }\href@noop {} {\bibfield  {journal} {\bibinfo  {journal} {Int. J.
  Mod. Phys. D}\ }\textbf {\bibinfo {volume} {20}},\ \bibinfo {pages} {413}
  (\bibinfo {year} {2011})},\ \Eprint {http://arxiv.org/abs/arXiv:1101.3864}
  {arXiv:1101.3864} \BibitemShut {NoStop}%
\bibitem [{\citenamefont {Koivisto}\ and\ \citenamefont
  {K.-Suonio}(2006)}]{palainst2}%
  \BibitemOpen
  \bibfield  {author} {\bibinfo {author} {\bibfnamefont {T.}~\bibnamefont
  {Koivisto}}\ and\ \bibinfo {author} {\bibfnamefont {H.}~\bibnamefont
  {K.-Suonio}},\ }\href@noop {} {\bibfield  {journal} {\bibinfo  {journal}
  {Class. Quant. Grav.}\ }\textbf {\bibinfo {volume} {23}},\ \bibinfo {pages}
  {2355} (\bibinfo {year} {2006})},\ \Eprint
  {http://arxiv.org/abs/astro-ph/0509422} {astro-ph/0509422} \BibitemShut
  {NoStop}%
\bibitem [{\citenamefont {Capozziello}\ \emph
  {et~al.}(2013{\natexlab{b}})\citenamefont {Capozziello}, \citenamefont
  {Harko}, \citenamefont {Koivisto}, \citenamefont {Lobo},\ and\ \citenamefont
  {Olmo}}]{hybridcosmo}%
  \BibitemOpen
  \bibfield  {author} {\bibinfo {author} {\bibfnamefont {S.}~\bibnamefont
  {Capozziello}}, \bibinfo {author} {\bibfnamefont {T.}~\bibnamefont {Harko}},
  \bibinfo {author} {\bibfnamefont {T.~S.}\ \bibnamefont {Koivisto}}, \bibinfo
  {author} {\bibfnamefont {F.~S.~N.}\ \bibnamefont {Lobo}}, \ and\ \bibinfo
  {author} {\bibfnamefont {G.~J.}\ \bibnamefont {Olmo}},\ }\href@noop {}
  {\bibfield  {journal} {\bibinfo  {journal} {JCAP}\ }\textbf {\bibinfo
  {volume} {1304}},\ \bibinfo {pages} {011} (\bibinfo {year}
  {2013}{\natexlab{b}})},\ \Eprint {http://arxiv.org/abs/arXiv:1209.2895v2}
  {arXiv:1209.2895v2} \BibitemShut {NoStop}%
\bibitem [{\citenamefont {Capozziello}\ \emph
  {et~al.}(2013{\natexlab{c}})\citenamefont {Capozziello}, \citenamefont
  {Harko}, \citenamefont {Koivisto}, \citenamefont {Lobo},\ and\ \citenamefont
  {Olmo}}]{hybridgala}%
  \BibitemOpen
  \bibfield  {author} {\bibinfo {author} {\bibfnamefont {S.}~\bibnamefont
  {Capozziello}}, \bibinfo {author} {\bibfnamefont {T.}~\bibnamefont {Harko}},
  \bibinfo {author} {\bibfnamefont {T.~S.}\ \bibnamefont {Koivisto}}, \bibinfo
  {author} {\bibfnamefont {F.~S.~N.}\ \bibnamefont {Lobo}}, \ and\ \bibinfo
  {author} {\bibfnamefont {G.~J.}\ \bibnamefont {Olmo}},\ }\href@noop {}
  {\bibfield  {journal} {\bibinfo  {journal} {JCAP}\ }\textbf {\bibinfo
  {volume} {07}},\ \bibinfo {pages} {024} (\bibinfo {year}
  {2013}{\natexlab{c}})},\ \Eprint {http://arxiv.org/abs/arXiv:1212.5817v2}
  {arXiv:1212.5817v2} \BibitemShut {NoStop}%
\bibitem [{\citenamefont {Lima}(2014)}]{hybridpert}%
  \BibitemOpen
  \bibfield  {author} {\bibinfo {author} {\bibfnamefont {N.~A.}\ \bibnamefont
  {Lima}},\ }\href@noop {} {\bibfield  {journal} {\bibinfo  {journal} {Phys.
  Rev. D}\ }\textbf {\bibinfo {volume} {89}},\ \bibinfo {pages} {083527}
  (\bibinfo {year} {2014})},\ \Eprint {http://arxiv.org/abs/arXiv:1402.4458}
  {arXiv:1402.4458} \BibitemShut {NoStop}%
\bibitem [{\citenamefont {Lima}\ and\ \citenamefont
  {S.-Barreto}(2016)}]{hybridback}%
  \BibitemOpen
  \bibfield  {author} {\bibinfo {author} {\bibfnamefont {N.~A.}\ \bibnamefont
  {Lima}}\ and\ \bibinfo {author} {\bibfnamefont {V.}~\bibnamefont
  {S.-Barreto}},\ }\href@noop {} {\bibfield  {journal} {\bibinfo  {journal}
  {Astroph. J.}\ }\textbf {\bibinfo {volume} {818}},\ \bibinfo {pages} {186}
  (\bibinfo {year} {2016})},\ \Eprint {http://arxiv.org/abs/arXiv:1501.05786}
  {arXiv:1501.05786} \BibitemShut {NoStop}%
\bibitem [{\citenamefont {Umezu}\ \emph {et~al.}(2005)\citenamefont {Umezu},
  \citenamefont {Ichiki},\ and\ \citenamefont {Yahiro}}]{mgc1}%
  \BibitemOpen
  \bibfield  {author} {\bibinfo {author} {\bibfnamefont {K.-i.}\ \bibnamefont
  {Umezu}}, \bibinfo {author} {\bibfnamefont {K.}~\bibnamefont {Ichiki}}, \
  and\ \bibinfo {author} {\bibfnamefont {M.}~\bibnamefont {Yahiro}},\ }\href
  {\doibase 10.1103/PhysRevD.72.044010} {\bibfield  {journal} {\bibinfo
  {journal} {Phys. Rev.}\ }\textbf {\bibinfo {volume} {D72}},\ \bibinfo {pages}
  {044010} (\bibinfo {year} {2005})},\ \Eprint
  {http://arxiv.org/abs/astro-ph/0503578} {arXiv:astro-ph/0503578 [astro-ph]}
  \BibitemShut {NoStop}%
%%CITATION = ASTRO-PH/0503578;%%
\bibitem [{\citenamefont {Galli}\ \emph {et~al.}(2009)\citenamefont {Galli},
  \citenamefont {Melchiorri}, \citenamefont {Smoot},\ and\ \citenamefont
  {Zahn}}]{mgc2}%
  \BibitemOpen
  \bibfield  {author} {\bibinfo {author} {\bibfnamefont {S.}~\bibnamefont
  {Galli}}, \bibinfo {author} {\bibfnamefont {A.}~\bibnamefont {Melchiorri}},
  \bibinfo {author} {\bibfnamefont {G.~F.}\ \bibnamefont {Smoot}}, \ and\
  \bibinfo {author} {\bibfnamefont {O.}~\bibnamefont {Zahn}},\ }\href {\doibase
  10.1103/PhysRevD.80.023508} {\bibfield  {journal} {\bibinfo  {journal} {Phys.
  Rev.}\ }\textbf {\bibinfo {volume} {D80}},\ \bibinfo {pages} {023508}
  (\bibinfo {year} {2009})},\ \Eprint {http://arxiv.org/abs/0905.1808}
  {arXiv:0905.1808 [astro-ph.CO]} \BibitemShut {NoStop}%
%%CITATION = ARXIV:0905.1808;%%
\bibitem [{\citenamefont {Lombriser}\ and\ \citenamefont
  {Taylor}(2015{\natexlab{a}})}]{lombriser:14b}%
  \BibitemOpen
  \bibfield  {author} {\bibinfo {author} {\bibfnamefont {L.}~\bibnamefont
  {Lombriser}}\ and\ \bibinfo {author} {\bibfnamefont {A.}~\bibnamefont
  {Taylor}},\ }\href {\doibase 10.1103/PhysRevLett.114.031101} {\bibfield
  {journal} {\bibinfo  {journal} {Phys. Rev. Lett.}\ }\textbf {\bibinfo
  {volume} {114}},\ \bibinfo {pages} {031101} (\bibinfo {year}
  {2015}{\natexlab{a}})},\ \Eprint {http://arxiv.org/abs/1405.2896}
  {arXiv:1405.2896 [astro-ph.CO]} \BibitemShut {NoStop}%
%%CITATION = ARXIV:1405.2896;%%
\bibitem [{\citenamefont {Lombriser}\ and\ \citenamefont
  {Taylor}(2015{\natexlab{b}})}]{lombriser:15a}%
  \BibitemOpen
  \bibfield  {author} {\bibinfo {author} {\bibfnamefont {L.}~\bibnamefont
  {Lombriser}}\ and\ \bibinfo {author} {\bibfnamefont {A.}~\bibnamefont
  {Taylor}},\ }\href {\doibase 10.1088/1475-7516/2015/11/040} {\bibfield
  {journal} {\bibinfo  {journal} {JCAP}\ }\textbf {\bibinfo {volume} {1511}},\
  \bibinfo {pages} {040} (\bibinfo {year} {2015}{\natexlab{b}})},\ \Eprint
  {http://arxiv.org/abs/1505.05915} {arXiv:1505.05915 [astro-ph.CO]}
  \BibitemShut {NoStop}%
%%CITATION = ARXIV:1505.05915;%%
\bibitem [{\citenamefont {Khoury}\ and\ \citenamefont
  {Weltman}(2004)}]{khoury:03}%
  \BibitemOpen
  \bibfield  {author} {\bibinfo {author} {\bibfnamefont {J.}~\bibnamefont
  {Khoury}}\ and\ \bibinfo {author} {\bibfnamefont {A.}~\bibnamefont
  {Weltman}},\ }\href {\doibase 10.1103/PhysRevLett.93.171104} {\bibfield
  {journal} {\bibinfo  {journal} {Phys. Rev. Lett.}\ }\textbf {\bibinfo
  {volume} {93}},\ \bibinfo {pages} {171104} (\bibinfo {year} {2004})},\
  \Eprint {http://arxiv.org/abs/astro-ph/0309300} {arXiv:astro-ph/0309300
  [astro-ph]} \BibitemShut {NoStop}%
%%CITATION = ASTRO-PH/0309300;%%
\bibitem [{\citenamefont {Hu}\ and\ \citenamefont {Sawicki}(2007)}]{hu:07}%
  \BibitemOpen
  \bibfield  {author} {\bibinfo {author} {\bibfnamefont {W.}~\bibnamefont
  {Hu}}\ and\ \bibinfo {author} {\bibfnamefont {I.}~\bibnamefont {Sawicki}},\
  }\href {\doibase 10.1103/PhysRevD.76.064004} {\bibfield  {journal} {\bibinfo
  {journal} {Phys. Rev.}\ }\textbf {\bibinfo {volume} {D76}},\ \bibinfo {pages}
  {064004} (\bibinfo {year} {2007})},\ \Eprint {http://arxiv.org/abs/0705.1158}
  {arXiv:0705.1158 [astro-ph]} \BibitemShut {NoStop}%
%%CITATION = ARXIV:0705.1158;%%
\bibitem [{\citenamefont {Brax}\ \emph {et~al.}(2008)\citenamefont {Brax},
  \citenamefont {van~de Bruck}, \citenamefont {Davis},\ and\ \citenamefont
  {Shaw}}]{brax:08}%
  \BibitemOpen
  \bibfield  {author} {\bibinfo {author} {\bibfnamefont {P.}~\bibnamefont
  {Brax}}, \bibinfo {author} {\bibfnamefont {C.}~\bibnamefont {van~de Bruck}},
  \bibinfo {author} {\bibfnamefont {A.-C.}\ \bibnamefont {Davis}}, \ and\
  \bibinfo {author} {\bibfnamefont {D.~J.}\ \bibnamefont {Shaw}},\ }\href
  {\doibase 10.1103/PhysRevD.78.104021} {\bibfield  {journal} {\bibinfo
  {journal} {Phys. Rev.}\ }\textbf {\bibinfo {volume} {D78}},\ \bibinfo {pages}
  {104021} (\bibinfo {year} {2008})},\ \Eprint {http://arxiv.org/abs/0806.3415}
  {arXiv:0806.3415 [astro-ph]} \BibitemShut {NoStop}%
%%CITATION = ARXIV:0806.3415;%%
\bibitem [{\citenamefont {Lombriser}(2014)}]{lombriser:14a}%
  \BibitemOpen
  \bibfield  {author} {\bibinfo {author} {\bibfnamefont {L.}~\bibnamefont
  {Lombriser}},\ }\href {\doibase 10.1002/andp.201400058} {\bibfield  {journal}
  {\bibinfo  {journal} {Annalen Phys.}\ }\textbf {\bibinfo {volume} {526}},\
  \bibinfo {pages} {259} (\bibinfo {year} {2014})},\ \Eprint
  {http://arxiv.org/abs/1403.4268} {arXiv:1403.4268 [astro-ph.CO]} \BibitemShut
  {NoStop}%
%%CITATION = ARXIV:1403.4268;%%
\bibitem [{\citenamefont {Bellini}\ and\ \citenamefont
  {Sawicki}(2014)}]{eftformalism}%
  \BibitemOpen
  \bibfield  {author} {\bibinfo {author} {\bibfnamefont {E.}~\bibnamefont
  {Bellini}}\ and\ \bibinfo {author} {\bibfnamefont {I.}~\bibnamefont
  {Sawicki}},\ }\href@noop {} {\bibfield  {journal} {\bibinfo  {journal}
  {JCAP}\ }\textbf {\bibinfo {volume} {07}},\ \bibinfo {pages} {050} (\bibinfo
  {year} {2014})},\ \Eprint {http://arxiv.org/abs/arXiv:1404.3713}
  {arXiv:1404.3713} \BibitemShut {NoStop}%
\bibitem [{\citenamefont {Sachs}\ and\ \citenamefont
  {Wolfe}(1967)}]{sachswolfe}%
  \BibitemOpen
  \bibfield  {author} {\bibinfo {author} {\bibfnamefont {R.~K.}\ \bibnamefont
  {Sachs}}\ and\ \bibinfo {author} {\bibfnamefont {A.~M.}\ \bibnamefont
  {Wolfe}},\ }\href@noop {} {\bibfield  {journal} {\bibinfo  {journal} {ApJ}\
  }\textbf {\bibinfo {volume} {147}},\ \bibinfo {pages} {73} (\bibinfo {year}
  {1967})}\BibitemShut {NoStop}%
\bibitem [{\citenamefont {Betoule}\ \emph {et~al.}(2014)\citenamefont {Betoule}
  \emph {et~al.}}]{Betoule}%
  \BibitemOpen
  \bibfield  {author} {\bibinfo {author} {\bibfnamefont {M.}~\bibnamefont
  {Betoule}} \emph {et~al.},\ }\href@noop {} {\bibfield  {journal} {\bibinfo
  {journal} {Astron. Astrophys.}\ }\textbf {\bibinfo {volume} {568}},\ \bibinfo
  {pages} {A22} (\bibinfo {year} {2014})},\ \Eprint
  {http://arxiv.org/abs/1401.4064} {1401.4064} \BibitemShut {NoStop}%
\bibitem [{\citenamefont {Riess}\ \emph {et~al.}(2011)\citenamefont {Riess}
  \emph {et~al.}}]{HST}%
  \BibitemOpen
  \bibfield  {author} {\bibinfo {author} {\bibfnamefont {A.~G.}\ \bibnamefont
  {Riess}} \emph {et~al.},\ }\href@noop {} {\bibfield  {journal} {\bibinfo
  {journal} {ApJ}\ }\textbf {\bibinfo {volume} {730}},\ \bibinfo {pages} {119}
  (\bibinfo {year} {2011})}\BibitemShut {NoStop}%
\bibitem [{\citenamefont {Beutler}\ \emph {et~al.}(2011)\citenamefont {Beutler}
  \emph {et~al.}}]{BAO1}%
  \BibitemOpen
  \bibfield  {author} {\bibinfo {author} {\bibfnamefont {F.}~\bibnamefont
  {Beutler}} \emph {et~al.},\ }\href@noop {} {\bibfield  {journal} {\bibinfo
  {journal} {Mon. Not. Roy. Astron. Soc.}\ }\textbf {\bibinfo {volume} {416}},\
  \bibinfo {pages} {3017} (\bibinfo {year} {2011})},\ \Eprint
  {http://arxiv.org/abs/arXiv:1106.3366} {arXiv:1106.3366} \BibitemShut
  {NoStop}%
\bibitem [{\citenamefont {Ross}\ \emph {et~al.}(2015)\citenamefont {Ross} \emph
  {et~al.}}]{BAO2}%
  \BibitemOpen
  \bibfield  {author} {\bibinfo {author} {\bibfnamefont {A.}~\bibnamefont
  {Ross}} \emph {et~al.},\ }\href@noop {} {\bibfield  {journal} {\bibinfo
  {journal} {Mon. Not. Roy. Astron. Soc.}\ }\textbf {\bibinfo {volume} {449}},\
  \bibinfo {pages} {835} (\bibinfo {year} {2015})},\ \Eprint
  {http://arxiv.org/abs/arXiv:1409.3242} {arXiv:1409.3242} \BibitemShut
  {NoStop}%
\bibitem [{\citenamefont {Anderson}\ \emph {et~al.}(2013)\citenamefont
  {Anderson} \emph {et~al.}}]{BAO3}%
  \BibitemOpen
  \bibfield  {author} {\bibinfo {author} {\bibfnamefont {L.}~\bibnamefont
  {Anderson}} \emph {et~al.},\ }\href@noop {} {\bibfield  {journal} {\bibinfo
  {journal} {MNRAS}\ }\textbf {\bibinfo {volume} {441}},\ \bibinfo {pages} {24}
  (\bibinfo {year} {2013})},\ \Eprint {http://arxiv.org/abs/arXiv:1312.4877}
  {arXiv:1312.4877} \BibitemShut {NoStop}%
\bibitem [{\citenamefont {collaboration}(2015)}]{Planck}%
  \BibitemOpen
  \bibfield  {author} {\bibinfo {author} {\bibfnamefont {P.}~\bibnamefont
  {collaboration}},\ }\href@noop {} {\bibfield  {journal} {\bibinfo  {journal}
  {ArXiv e-prints}\ ,\ \bibinfo {pages} {40}} (\bibinfo {year} {2015})},\
  \Eprint {http://arxiv.org/abs/arXiv:1502.01582} {arXiv:1502.01582}
  \BibitemShut {NoStop}%
\bibitem [{\citenamefont {Lewis}\ and\ \citenamefont {Bridle}(2002)}]{cosmomc}%
  \BibitemOpen
  \bibfield  {author} {\bibinfo {author} {\bibfnamefont {A.}~\bibnamefont
  {Lewis}}\ and\ \bibinfo {author} {\bibfnamefont {S.}~\bibnamefont {Bridle}},\
  }\href {\doibase 10.1103/PhysRevD.66.103511} {\bibfield  {journal} {\bibinfo
  {journal} {Phys. Rev.}\ }\textbf {\bibinfo {volume} {D66}},\ \bibinfo {pages}
  {103511} (\bibinfo {year} {2002})},\ \Eprint
  {http://arxiv.org/abs/astro-ph/0205436} {arXiv:astro-ph/0205436 [astro-ph]}
  \BibitemShut {NoStop}%
%%CITATION = ASTRO-PH/0205436;%%
\bibitem [{\citenamefont {Calabrese}\ \emph {et~al.}(2009)\citenamefont
  {Calabrese} \emph {et~al.}}]{cmblens}%
  \BibitemOpen
  \bibfield  {author} {\bibinfo {author} {\bibfnamefont {E.}~\bibnamefont
  {Calabrese}} \emph {et~al.},\ }\href@noop {} {\bibfield  {journal} {\bibinfo
  {journal} {Phys. Rev. D}\ }\textbf {\bibinfo {volume} {80}},\ \bibinfo
  {pages} {103516} (\bibinfo {year} {2009})},\ \Eprint
  {http://arxiv.org/abs/arXiv:0908.1585v1} {arXiv:0908.1585v1} \BibitemShut
  {NoStop}%
\bibitem [{\citenamefont {{Madau}}\ \emph {et~al.}(1997)\citenamefont
  {{Madau}}, \citenamefont {{Meiksin}},\ and\ \citenamefont
  {{Rees}}}]{madau:97}%
  \BibitemOpen
  \bibfield  {author} {\bibinfo {author} {\bibfnamefont {P.}~\bibnamefont
  {{Madau}}}, \bibinfo {author} {\bibfnamefont {A.}~\bibnamefont {{Meiksin}}},
  \ and\ \bibinfo {author} {\bibfnamefont {M.~J.}\ \bibnamefont {{Rees}}},\
  }\href@noop {} {\bibfield  {journal} {\bibinfo  {journal} {\apj}\ }\textbf
  {\bibinfo {volume} {475}},\ \bibinfo {pages} {429} (\bibinfo {year}
  {1997})},\ \Eprint {http://arxiv.org/abs/astro-ph/9608010} {astro-ph/9608010}
  \BibitemShut {NoStop}%
\bibitem [{\citenamefont {Brax}\ \emph {et~al.}(2013)\citenamefont {Brax},
  \citenamefont {Clesse},\ and\ \citenamefont {Davis}}]{brax:12}%
  \BibitemOpen
  \bibfield  {author} {\bibinfo {author} {\bibfnamefont {P.}~\bibnamefont
  {Brax}}, \bibinfo {author} {\bibfnamefont {S.}~\bibnamefont {Clesse}}, \ and\
  \bibinfo {author} {\bibfnamefont {A.-C.}\ \bibnamefont {Davis}},\ }\href
  {\doibase 10.1088/1475-7516/2013/01/003} {\bibfield  {journal} {\bibinfo
  {journal} {JCAP}\ }\textbf {\bibinfo {volume} {1301}},\ \bibinfo {pages}
  {003} (\bibinfo {year} {2013})},\ \Eprint {http://arxiv.org/abs/1207.1273}
  {arXiv:1207.1273 [astro-ph.CO]} \BibitemShut {NoStop}%
%%CITATION = ARXIV:1207.1273;%%
\bibitem [{\citenamefont {Hall}\ \emph {et~al.}(2013)\citenamefont {Hall},
  \citenamefont {Bonvin},\ and\ \citenamefont {Challinor}}]{hall:12}%
  \BibitemOpen
  \bibfield  {author} {\bibinfo {author} {\bibfnamefont {A.}~\bibnamefont
  {Hall}}, \bibinfo {author} {\bibfnamefont {C.}~\bibnamefont {Bonvin}}, \ and\
  \bibinfo {author} {\bibfnamefont {A.}~\bibnamefont {Challinor}},\ }\href
  {\doibase 10.1103/PhysRevD.87.064026} {\bibfield  {journal} {\bibinfo
  {journal} {Phys. Rev.}\ }\textbf {\bibinfo {volume} {D87}},\ \bibinfo {pages}
  {064026} (\bibinfo {year} {2013})},\ \Eprint {http://arxiv.org/abs/1212.0728}
  {arXiv:1212.0728 [astro-ph.CO]} \BibitemShut {NoStop}%
%%CITATION = ARXIV:1212.0728;%%
\bibitem [{\citenamefont {Schutz}(1986)}]{schutz:86}%
  \BibitemOpen
  \bibfield  {author} {\bibinfo {author} {\bibfnamefont {B.~F.}\ \bibnamefont
  {Schutz}},\ }\href {\doibase 10.1038/323310a0} {\bibfield  {journal}
  {\bibinfo  {journal} {Nature}\ }\textbf {\bibinfo {volume} {323}},\ \bibinfo
  {pages} {310} (\bibinfo {year} {1986})}\BibitemShut {NoStop}%
%%CITATION = NATUA,323,310;%%
\bibitem [{\citenamefont {Holz}\ and\ \citenamefont {Hughes}(2005)}]{holz:05}%
  \BibitemOpen
  \bibfield  {author} {\bibinfo {author} {\bibfnamefont {D.~E.}\ \bibnamefont
  {Holz}}\ and\ \bibinfo {author} {\bibfnamefont {S.~A.}\ \bibnamefont
  {Hughes}},\ }\href {\doibase 10.1086/431341} {\bibfield  {journal} {\bibinfo
  {journal} {Astrophys. J.}\ }\textbf {\bibinfo {volume} {629}},\ \bibinfo
  {pages} {15} (\bibinfo {year} {2005})},\ \Eprint
  {http://arxiv.org/abs/astro-ph/0504616} {arXiv:astro-ph/0504616 [astro-ph]}
  \BibitemShut {NoStop}%
%%CITATION = ASTRO-PH/0504616;%%
\bibitem [{\citenamefont {Cutler}\ and\ \citenamefont
  {Holz}(2009)}]{cutler:09}%
  \BibitemOpen
  \bibfield  {author} {\bibinfo {author} {\bibfnamefont {C.}~\bibnamefont
  {Cutler}}\ and\ \bibinfo {author} {\bibfnamefont {D.~E.}\ \bibnamefont
  {Holz}},\ }\href {\doibase 10.1103/PhysRevD.80.104009} {\bibfield  {journal}
  {\bibinfo  {journal} {Phys. Rev.}\ }\textbf {\bibinfo {volume} {D80}},\
  \bibinfo {pages} {104009} (\bibinfo {year} {2009})},\ \Eprint
  {http://arxiv.org/abs/0906.3752} {arXiv:0906.3752 [astro-ph.CO]} \BibitemShut
  {NoStop}%
%%CITATION = ARXIV:0906.3752;%%
\bibitem [{\citenamefont {Tamanini}\ \emph {et~al.}(2016)\citenamefont
  {Tamanini}, \citenamefont {Caprini}, \citenamefont {Barausse}, \citenamefont
  {Sesana}, \citenamefont {Klein},\ and\ \citenamefont
  {Petiteau}}]{tamanini:16}%
  \BibitemOpen
  \bibfield  {author} {\bibinfo {author} {\bibfnamefont {N.}~\bibnamefont
  {Tamanini}}, \bibinfo {author} {\bibfnamefont {C.}~\bibnamefont {Caprini}},
  \bibinfo {author} {\bibfnamefont {E.}~\bibnamefont {Barausse}}, \bibinfo
  {author} {\bibfnamefont {A.}~\bibnamefont {Sesana}}, \bibinfo {author}
  {\bibfnamefont {A.}~\bibnamefont {Klein}}, \ and\ \bibinfo {author}
  {\bibfnamefont {A.}~\bibnamefont {Petiteau}},\ }\href@noop {} {\  (\bibinfo
  {year} {2016})},\ \Eprint {http://arxiv.org/abs/1601.07112} {arXiv:1601.07112
  [astro-ph.CO]} \BibitemShut {NoStop}%
%%CITATION = ARXIV:1601.07112;%%
\bibitem [{\citenamefont {Hojjati}\ \emph {et~al.}(2011)\citenamefont
  {Hojjati}, \citenamefont {Pogosian},\ and\ \citenamefont {Zhao}}]{mgcamb}%
  \BibitemOpen
  \bibfield  {author} {\bibinfo {author} {\bibfnamefont {A.}~\bibnamefont
  {Hojjati}}, \bibinfo {author} {\bibfnamefont {L.}~\bibnamefont {Pogosian}}, \
  and\ \bibinfo {author} {\bibfnamefont {G.-B.}\ \bibnamefont {Zhao}},\
  }\href@noop {} {\bibfield  {journal} {\bibinfo  {journal} {JCAP}\ }\textbf
  {\bibinfo {volume} {08}},\ \bibinfo {pages} {005} (\bibinfo {year} {2011})},\
  \Eprint {http://arxiv.org/abs/arXiv:1106.4543} {arXiv:1106.4543} \BibitemShut
  {NoStop}%
\bibitem [{\citenamefont {Lewis}\ \emph {et~al.}(2000)\citenamefont {Lewis},
  \citenamefont {Challinor},\ and\ \citenamefont {Lasenby}}]{camb}%
  \BibitemOpen
  \bibfield  {author} {\bibinfo {author} {\bibfnamefont {A.}~\bibnamefont
  {Lewis}}, \bibinfo {author} {\bibfnamefont {A.}~\bibnamefont {Challinor}}, \
  and\ \bibinfo {author} {\bibfnamefont {A.}~\bibnamefont {Lasenby}},\ }\href
  {http://camb.info/} {\bibfield  {journal} {\bibinfo  {journal} {Astroph. J.}\
  }\textbf {\bibinfo {volume} {538}},\ \bibinfo {pages} {473} (\bibinfo {year}
  {2000})},\ \Eprint {http://arxiv.org/abs/astro-ph/9911177} {astro-ph/9911177}
  \BibitemShut {NoStop}%
\end{thebibliography}%

\appendix

\section{Implementation in {\sc mgcamb}} \label{phenom}

In order to compute the CMB observables, we implement our early decaying modified gravity model in the publicly available {\sc mgcamb} code~\cite{mgcamb}, a modified version of the also public {\sc camb} code~\cite{camb} that allows to study the effects of modified gravity models on the CMB through modifications of the linear equations describing the growth of perturbations.
{\sc mgcamb} works by parameterizing the evolution of the gravitational potentials simply through two time- and scale-dependent functions: the ratio of the metric potentials $\gamma(a,k) \equiv \Phi/\Psi$ and the effective modified gravitational coupling in the Poisson equation, $\mu(a,k)=G_{\rm eff}/G$.
The framework of {\sc mgcamb} is general enough to include possible early-time effects, hence it is well-suited for testing the hybrid metric-Palatini theory.
Moreover, we chose to work with {\sc mgcamb} as it allows us to use the approximations described in Secs.~\ref{subapp} and \ref{earlyapp} to improve computational efficiency without loss of accuracy.

\begin{figure}
 \includegraphics[scale=0.45]{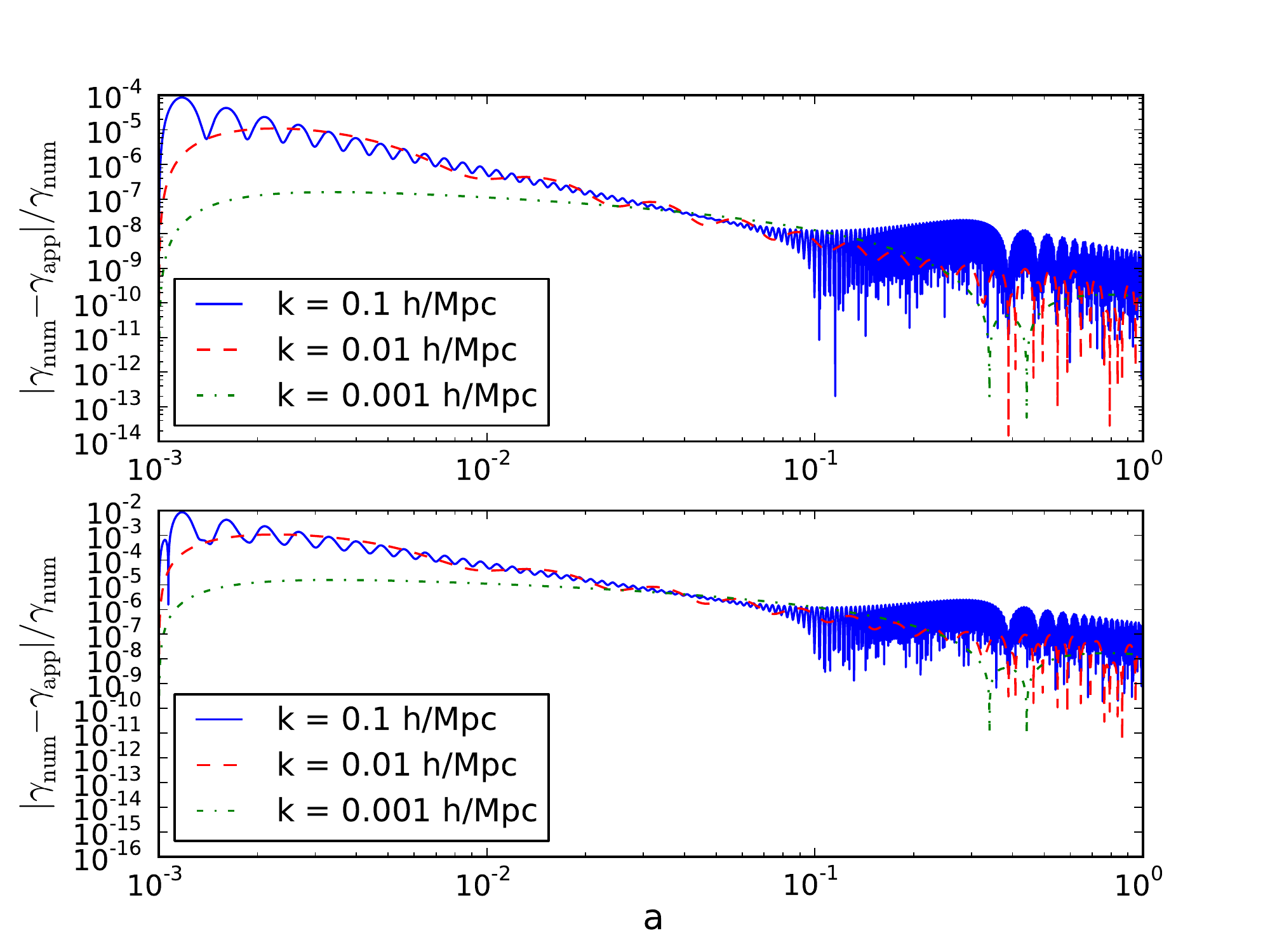}
 \caption{\label{ratio_app} Relative difference between the numerical evolution of $\gamma \equiv \Phi/\Psi$ and the approximation in Eq.~(\ref{mgcambratio}).
The top panel shows $|f_{\mathcal{R}{\rm{i}}}| = 10^{-4}$ and the lower panel shows $|f_{\mathcal{R}{\rm{i}}}| = 10^{-2}$. We have again fixed $\Omega_{\rm{m}} = 0.30$.}
\end{figure}

We implement our model by modifying both $\gamma$ and $\mu$ in the code.
For $\gamma$ we use the subhorizon approximation described in Eq.~(\ref{qsratio}) and add an oscillatory term described by $\delta f_{\mathcal{R}}$ to account for the early-time oscillations.
From Ref.~\cite{hybridpert} we note that the gravitational potentials can be expressed as
\begin{equation}{\label{potentials_mgcamb}}
 \Phi = \Phi_{+} + \frac{\delta f_{\mathcal{R}}}{2(1+f_{\mathcal{R}})} \,, \hspace{2 mm} \Psi = \Phi_{+} - \frac{\delta f_{\mathcal{R}}}{2(1+f_{\mathcal{R}})} \,,
\end{equation}
which uses the observation that the early-time oscillations in $\delta f_{\mathcal{R}}$ do not affect the lensing potential $\Phi_{+}$ for small-enough values of the amplitude of the oscillations.
$\Phi_{+}$ has an approximately constant value of unity throughout the matter dominated era. Therefore, with $\Phi_{+} \gg \delta f_{\mathcal{R}}$ one can perform a Taylor expansion on the ratio between the potentials that results in
\begin{equation}{\label{mgcambratio}}
 \gamma = \frac{\Phi}{\Psi} \approx 1 - \frac{\delta f_{\mathcal{R}}}{(1+f_{\mathcal{R}})} \,.
\end{equation}
We compare this approximation against numerical results in Fig.~\ref{ratio_app}, finding good agreement between the two, at an accuracy comparable to that observed in Fig.~\ref{delphicomp} for the slip between the metric potentials.
Given this result, we generalize $\gamma_{\rm{QS}}$ with the simple modification
\begin{equation}{\label{gamma_mgcamb}}
 \gamma_{\rm{MG}} \approx \gamma_{\rm{QS}} + \frac{\delta f_{\mathcal{R}}}{1+f_{\mathcal{R}}},
\end{equation}
where $\gamma_{\rm{QS}}$ can be found in Eq.~(\ref{qsratio}).
Correspondingly, we modify $\mu$ to include the effect of the oscillations in the Poisson equation such that
\begin{equation}{\label{mu_mgcamb}}
 \mu_{\rm{MG}} = \mu_{\rm{QS}} + \frac{\delta f_{\mathcal{R}}}{2 (1+f_{\mathcal{R}})} \,,
\end{equation}
where $\mu_{\rm{QS}}$ is given in Eq.~(\ref{psiqs2}).

Finally, note that the initial conditions required to solve for the background evolution of our models are always set at the redshift $z_{\rm{i}} = 1000$.
As described in Secs.~\ref{sec:decoupling} and \ref{coneft} through an embedding in the effective field theory of Horndeski gravity, the model is designed to behave as $\Lambda$CDM at the level of linear perturbations down to a redshift $z_{\rm{on}}$, at which point the modifications are introduced.
At redshift $z_{\rm{i}}$ we set $\delta f_{\mathcal{R}} = 0$, with its subsequent evolution being determined by Eq.~(\ref{delphiwkbapp}).

\section{Analytic Solution for the Integrated Spring Term} \label{appendix1}

Using Eq.~(\ref{ratiophipphi}), we can simplify the $w$ term of Eq.~(\ref{delphiwkbapp}) as
\begin{equation}{\label{wapproximation}}
w \approx \PR{\frac{k^2 a^{-2}}{ H_{0}^{2} E} + \frac{\PC{a_{\rm{aux}}-\sqrt{d}}^{2}}{2} + \frac{\Omega_{\rm{m}}}{\Omega_{\rm{m}}+\Omega_{\Lambda}a^{3}}}^{1/2} \,,
\end{equation}
where we have neglected the presence of radiation in the Hubble factor $H \equiv H_0 \sqrt{E}$ since applying this approximation only for redshifts deep within the matter-dominated era.
For $k \gg a H$, Eq.~(\ref{wapproximation}) can be further approximated by
\begin{equation}{\label{wapplargek}}
w \approx \PC{\frac{k^2}{a^2 H_0^{2}E}}^{1/2}\PC{1 + \frac{b a^2 H_{0}^{2}E}{2 k^2}} \,, 
\end{equation}
where $b = \PC{a_{\rm{aux}}-\sqrt{d}}^{2}/2 + \Omega_{\rm{m}}/\PC{\Omega_{\rm{m}}+\Omega_{\Lambda}a^3}$,
which allows us to perform an analytic integration of Eq.~(\ref{delphiwkbapp}).
The result depends on hypergeometric functions that can, however, be approximated as unity.
For simplicity, we therefore present the result without the presence of these functions:
\begin{widetext}
\begin{equation}{\label{intlargek}}
\int w d \ln a  \approx 2 \PC{\frac{k^2 a}{H_{0}^2 \Omega_{\rm{m}}}}^{1/2} + \frac{\PR{a_{\rm{aux}}-\sqrt{d}}^{2}}{4}\PC{\frac{\Omega_{\rm{m}}H_{0}^{2}}{k^2 a}}^{1/2}\PC{\sqrt{\frac{\Omega_{\rm{m}}}{\Omega_{\Lambda}}a^3+1}-3} - \Omega_{\rm{m}}\PC{\frac{H_{0}^{2}}{k^2 a}}^{1/2}.
\end{equation}
\end{widetext}

In the limit of $k \ll a H$, we can instead approximate $w$ as
\begin{equation}{\label{wappsamllk}}
w \approx \sqrt{b}\PC{1 + \frac{1}{2}\frac{k^2}{a^2 H_{0}^{2} E b}} \,.
\end{equation}
To perform an analytic integration, we use the approximation $b \approx \PC{a_{\rm{aux}}-\sqrt{d}}^{2}/2 + 1$, which results in
\begin{equation}{\label{intsmallk}}
\int w d \ln a \approx \sqrt{b} + \frac{1}{2 \sqrt{b}} \frac{k^2 a}{H_{0}^{2} \Omega_{\rm{m}}} \,.
\end{equation}

We compare the implementation of the approximations in Eqs.~(\ref
{wappsamllk}) and (\ref{intsmallk}) against numerical results in Fig.~\ref{delphicomp}, finding good agreement between the two.

\end{document}